\documentclass[twocolumn]{aastex631}
\usepackage{multirow}
\usepackage[T1]{fontenc}
\usepackage[utf8]{inputenc}
\usepackage{appendix}
\usepackage{amsmath}

\DeclareUnicodeCharacter{02BC}{'}
\begin{document}

\title{Expanding the ultracompacts: gravitational wave-driven mass transfer in the shortest-period binaries with accretion disks}

\author[0000-0002-0568-6000]{Joheen Chakraborty}
\author[0000-0002-7226-836X]{Kevin B. Burdge}
\author[0000-0003-3182-5569]{Saul A. Rappaport}
\affiliation{Department of Physics \& Kavli Institute for Astrophysics and Space Research, Massachusetts Institute of Technology,\\Cambridge, MA 02139, USA}
\author{James Munday}
\affiliation{Department of Physics, Gibbet Hill Road, University of Warwick, Coventry CV4 7AL, United Kingdom}
\author{Hai-Liang Chen}
\affiliation{Yunnan Observatories, Chinese Academy of Sciences (CAS), Kunming 650216, People’s Republic of China}
\affiliation{International Centre of Supernovae, Yunnan Key Laboratory, Kunming 650216, Peopleʼs Republic of China}
\author[0000-0002-4717-5102]{Pablo Rodríguez-Gil}
\affiliation{Instituto de Astrofísica de Canarias, E-38205 La Laguna, Tenerife, Spain}
\affiliation{Departamento de Astrofísica, Universidad de La Laguna, E-38206 La Laguna, Tenerife, Spain}
\author[0000-0003-4236-9642]{V. S. Dhillon}
\affiliation{Astrophysics Research Cluster, School of Mathematical and Physical Sciences, University of Sheffield, Sheffield S3 7RH, UK}
\affiliation{Instituto de Astrofísica de Canarias, E-38205 La Laguna, Tenerife, Spain}
\author[0000-0001-6211-1388]{Scott A. Hughes}
\affiliation{Department of Physics \& Kavli Institute for Astrophysics and Space Research, Massachusetts Institute of Technology,\\Cambridge, MA 02139, USA}
\author[0000-0002-0752-2974]{Gijs Nelemans}
\affiliation{Department of Astrophysics/IMAPP, Radboud University, P.O. Box 9010, NL-6500 GL Nijmegen, The Netherlands}
\affiliation{Institute of Astronomy, KU Leuven, Celestijnenlaan 200D, B-3001 Leuven, Belgium}
\affiliation{SRON, Netherlands Institute for Space Research, Sorbonnelaan 2, NL-3584 CA Utrecht, The Netherlands}
\author[0000-0003-0172-0854]{Erin Kara}
\affiliation{Department of Physics \& Kavli Institute for Astrophysics and Space Research, Massachusetts Institute of Technology,\\Cambridge, MA 02139, USA}
\author[0000-0001-8018-5348]{Eric C. Bellm}
\affiliation{DIRAC Institute, Department of Astronomy, University of Washington, 3910 15th Avenue NE, Seattle, WA 98195, USA}
\author[0000-0002-3316-7240]{Alex J. Brown}
\affiliation{Departament de Física, Universitat Politècnica de Catalunya, c/Esteve Terrades 5, 08860 Castelldefels, Spain}
\author[0000-0002-5870-0443]{Noel Castro Segura}
\affiliation{Department of Physics, Gibbet Hill Road, University of Warwick, Coventry CV4 7AL, United Kingdom}
\author[0000-0001-9152-6224]{Tracy X. Chen}
\affiliation{IPAC, California Institute of Technology, 1200 E. California Blvd, Pasadena, CA 91125, USA}
\author[0000-0003-4780-4105]{Emma Chickles}
\affiliation{Department of Physics \& Kavli Institute for Astrophysics and Space Research, Massachusetts Institute of Technology,\\Cambridge, MA 02139, USA}
\author[0000-0003-3665-5482]{Martin J. Dyer}
\affiliation{Astrophysics Research Cluster, School of Mathematical and Physical Sciences, University of Sheffield, Sheffield S3 7RH, UK}
\author[0000-0002-5884-7867]{Richard Dekany}
\affiliation{Caltech Optical Observatories, California Institute of Technology, Pasadena, CA 91125, USA}
\author{Andrew J. Drake}
\affiliation{Division of Physics, Mathematics \&\ Astronomy, California Institute of Technology, Pasadena, CA 91125, USA}
\author[0009-0007-5535-3312]{James Garbutt}
\affiliation{Astrophysics Research Cluster, School of Mathematical and Physical Sciences, University of Sheffield, Sheffield S3 7RH, UK}
\author[0000-0002-3168-0139]{Matthew J. Graham}
\affiliation{Division of Physics, Mathematics \&\ Astronomy, California Institute of Technology, Pasadena, CA 91125, USA}
\author[0000-0002-0948-4801]{Matthew J. Green}
\affiliation{Max-Planck-Institut für Astronomie, Königstuhl 17, D-69117 Heidelberg, Germany}
\author[0009-0004-3067-2227]{Dan Jarvis}
\affiliation{Astrophysics Research Cluster, School of Mathematical and Physical Sciences, University of Sheffield, Sheffield S3 7RH, UK}
\author[0000-0001-6894-6044]{Mark R. Kennedy}
\affiliation{School of Physics, University College Cork, Cork, Ireland}
\author{Paul Kerry}
\affiliation{Astrophysics Research Cluster, School of Mathematical and Physical Sciences, University of Sheffield, Sheffield S3 7RH, UK}
\author[0000-0001-5390-8563]{S.\ R.\ Kulkarni}
\affiliation{Division of Physics, Mathematics \&\ Astronomy, California Institute of Technology, Pasadena, CA 91125, USA}
\author[0000-0001-7221-855X]{Stuart P. Littlefair}
\affiliation{Astrophysics Research Cluster, School of Mathematical and Physical Sciences, University of Sheffield, Sheffield S3 7RH, UK}
\author[0000-0003-2242-0244]{Ashish A. Mahabal}
\affiliation{Division of Physics, Mathematics \&\ Astronomy, California Institute of Technology, Pasadena, CA 91125, USA}
\author[0000-0002-8532-9395]{Frank J. Masci}
\affiliation{IPAC, California Institute of Technology, 1200 E. California Blvd, Pasadena, CA 91125, USA}
\author[0000-0003-1631-4170]{James McCormac}
\affiliation{Department of Physics, Gibbet Hill Road, University of Warwick, Coventry CV4 7AL, United Kingdom}
\author[0000-0002-2695-2654]{Steven G. Parsons}
\affiliation{Astrophysics Research Cluster, School of Mathematical and Physical Sciences, University of Sheffield, Sheffield S3 7RH, UK}
\author[0000-0003-4615-6556]{Ingrid Pelisoli}
\affiliation{Department of Physics, Gibbet Hill Road, University of Warwick, Coventry CV4 7AL, United Kingdom}
\author{Eleanor Pike}
\affiliation{Astrophysics Research Cluster, School of Mathematical and Physical Sciences, University of Sheffield, Sheffield S3 7RH, UK}
\author{Thomas A. Prince}
\affiliation{Division of Physics, Mathematics \&\ Astronomy, California Institute of Technology, Pasadena, CA 91125, USA}
\author[0000-0002-0387-370X]{Reed Riddle}
\affiliation{Division of Physics, Mathematics \&\ Astronomy, California Institute of Technology, Pasadena, CA 91125, USA}
\author[0000-0002-2626-2872]{Jan van Roestel}
\affiliation{Anton Pannekoek Institute for Astronomy, University of Amsterdam, 1090 GE Amsterdam, The Netherlands}
\author{Dave Sahman}
\affiliation{Astrophysics Research Cluster, School of Mathematical and Physical Sciences, University of Sheffield, Sheffield S3 7RH, UK}
\author[0000-0002-9998-6732]{Avery Wold}
\affiliation{IPAC, California Institute of Technology, 1200 E. California Blvd, Pasadena, CA 91125, USA}
\author[0000-0001-9195-7390]{Tin Long Sunny Wong}
\affiliation{Department of Physics, University of California, Santa Barbara, CA 93106, USA}

\begin{abstract}

We report the discovery of three ultracompact binary white dwarf systems hosting accretion disks, with orbital periods of 7.95, 8.68, and 13.15 minutes. This significantly augments the population of mass-transferring binaries at the shortest periods, and provides the first evidence that accretors in ultracompacts can be dense enough to host accretion disks even below 10 minutes (where previously only direct-impact accretors were known). In the two shortest-period systems, we measured changes in the orbital periods driven by the combined effect of gravitational wave emission and mass transfer; we find $\dot{P}$ is negative in one case, and positive in the other. This is only the second system measured with a positive $\dot{P}$, and it the most compact binary known that has survived a period minimum. Using these systems as examples, we show how the measurement of $\dot{P}$ is a powerful tool in constraining the physical properties of binaries, e.g. the mass and mass-radius relation of the donor stars. We find that the chirp masses of ultracompact binaries at these periods seem to cluster around $\mathcal{M}_c \sim 0.3 M_\odot$, perhaps suggesting a common origin for these systems or a selection bias in electromagnetic discoveries. Our new systems are among the highest-amplitude known gravitational wave sources in the millihertz regime, providing exquisite opportunity for multi-messenger study with future space-based observatories such as \textit{LISA} and TianQin; we discuss how such systems provide fascinating laboratories to study the unique regime where the accretion process is mediated by gravitational waves.
\end{abstract}
\date{Accepted 2024 November 18. Received 2024 November 15; in original form 2024 October 11.}

\section{Introduction} \label{sec:intro}

Binary systems containing accreting white dwarfs (WDs) with orbital periods $\lesssim 65$ minutes---the minimum period at which a hydrogen-rich star can attain the density to fit within its Roche lobe---are known as AM Canum Venaticorum (AM CVn) systems (for reviews see \citealt{Nelemans2005,Solheim2010,Ramsay2018}), which are part of the class of ultracompact binaries. The dense donor stars in AM CVns lack hydrogen, and the systems show spectra comprising stellar material dominated by helium, carbon, and nitrogen \citep{Warner1995,Nelemans2010}. 

Several dozen AM CVns are known, discovered via an eclectic range of signatures including optical outbursts \citep{Levitan2015}, X-ray pulsations \citep{Israel1999}, spectral/color properties \citep{Roelofs2007,Roelofs2009,Rau2010,Rodriguez2023,Rodriguez2024}, and short-period photometric variability in the accretion disks \citep{Smak1967,Green2018b,Burdge2020a,vanRoestel2022}. The current population has orbital periods spanning from 65 down to 5 minutes \citep{Ramsay2018,Green2024,Kupfer2024}. Below 10 minutes, all known systems are so compact that the accretion stream directly collides with the accretor surface (direct-impact accretion, \citealt{Marsh2002,Marsh2004}).

The extremely short periods of these objects mean emission of gravitational waves (GWs) is the primary factor driving their orbital evolution \citep{Paczynski1967}. They will be the dominant source of millihertz gravitational waves detected by future space-baced observatories, such as the Laser Space Antenna Interferometer \citep[\textit{LISA};][]{Amaro-Seoane2017,Amaro-Seoane2023}, TianQin \citep{Luo2016}, and Taiji \citep{Ruan2020} missions. As the instantaneous GW strain amplitude, expressed in terms of the chirp mass ($\mathcal{M}_c$), orbital period ($P$), and distance ($d$), is roughly \citep{Nelemans2004}:
\begin{equation}
h = 0.5 \times 10^{-21} \bigg(\frac{\mathcal{M}_c}{\mathrm{M}_\odot}\bigg)^{5/3} \bigg(\frac{P}{\mathrm{1\;h}}\bigg)^{-2/3} \bigg(\frac{d}{\mathrm{1\;kpc}}\bigg)^{-1}~,
\end{equation}
the shortest-period systems will typically have the highest intrinsic strains and most rapid evolution. Having inspiral set primarily by GWs makes AM CVns unique among accreting binary systems; the certainty of our understanding of gravitational radiation provides a convenient foothold in understanding their complex accretion properties, in contrast to binaries with poorly-understood angular momentum loss mechanisms (e.g. magnetic braking).

Apart from their nature as GW sources, ultracompact binaries provide interesting laboratories for astrophysics. They are potential progenitors of Type Ia supernovae, making them central to understanding one of the foundational standard candles for local-universe cosmology both observationally \citep{Maoz2014,Jha2019} and theoretically \citep{Hillebrandt2013,Shen2014,Liu2023}. Yet we lack a complete understanding of AM CVn formation and evolution, which has significant implications for their effect on SNIa rates and, more generally, the long-term fate of binary stars in the Galaxy. There are up to three channels contributing significantly to the population, most easily distinguished via their imprints on the donor C/O, N/O, and N/C ratios \citep{Nelemans2010,Green2018a}:
\begin{enumerate}
    \item \textbf{Double white dwarf channel:} two detached white dwarfs evolve closer due to gravitational wave emission following a common envelope phase, eventually starting mass transfer at $P\lesssim 10$ min \citep{Nelemans2001}. Depending whether tidal synchronization efficiently returns angular momentum to the donor, their orbits \textit{may} stabilize, reaching a period minimum at 4-9 minutes before outspiraling to produce a long-lived AM CVn \citep{Marsh2004,Kaplan2012}. The donors are He-dominated, with a possible low-mass residual hydrogen envelope. They have metal abundances according to the equilibrium of the CNO cycle depending on the main sequence progenitor mass. In the relevant mass ranges for AM CVn donors, they transfer mass with equilibrium N/C $\approx 100$.
    \item \textbf{Helium star channel:} If the donor is instead a helium core-burning star (e.g. an sdB/O star, \citealt{Heber2016}), a minimum period of $P \approx 11$ minutes is reached before the onset of mass transfer \citep{Savonije1986}. The orbital evolution is set by the degeneracy level in the mass-transferring outer envelope. N/O and N/C $\ll 100$ are expected, because fusion and CNO processing were interrupted to produce these donors \citep{Yungelson2008}.
    \item \textbf{Hydrogen CV channel:} For a partially evolved main sequence donor, mass transfer may begin while the system is still hydrogen-rich. The stripped donor gradually becomes degenerate enough to reach a compact orbit. These systems will have undergone H fusion the longest, so N/C will be $>100$ \citep{Kalomeni2016}, and will also show traces of H. It is uncertain how significantly this channel contributes to the shortest-period population, and our understanding of their period minima is still evolving \citep{Ramsay2018,Burdge2022}.
\end{enumerate}
Discovery and spectral characterization of further AM CVns is the most promising route to determining the importance of each channel, which then constitutes a key input into binary population synthesis models \citep{Breivik2020}, the all-sky millihertz GW signal \citep{Littenberg2020}, and Type Ia SNa rates \citep{Ruiter2009}.

Here we report the discovery of three new ultracompact binaries with accretion disks: ZTF J0546+3843, ZTF J1858--2024, and ZTF J0425+3858. Two of these systems have orbital periods under 10 minutes, where previously only direct-impact accretors were known. Our systems are thus the most compact known binaries with accretion disks, showing that the accretors in AM CVns can become dense enough to host disks even at $P<10$~min. Our study significantly augments the population for systematic study of binary evolution closest to the period minimum, and provides exquisite laboratories to test the accretion process mediated by strong gravitational radiation.

In Section~\ref{sec:methods} we describe the observations and data analysis procedures used in this work. In Section~\ref{sec:results} we give an overview of the results, including time-resolved photometry and spectroscopy and phase-coherent timing. In Section~\ref{sec:discussion} we discuss implications of the spectroscopic and timing measurements of these systems on AM CVn evolution. We further describe how direct measurement of a period derivative is a powerful tool for significantly constraining binary systems driven by GWs, and that their high GW strain will aid in fully characterizing them with the future space-based GW detectors. We make remarks on the thus-far observed population of binaries below $\sim 15$ minutes. In Section~\ref{sec:conclusion} we make concluding remarks and suggest directions for future wo
\section{Methods and Observations} \label{sec:methods}
\begin{table*}
\centering
\begin{tabular}{c|c|c|c|c|c|c}
Name & $P_0$ & RA (J2000.0) & Dec (J2000.0) & Distance (est.) &  Optical mag. ($L_{\mathrm{opt}}$) & $F_X$ ($L_X$) \\
 & (min) & (deg) & (deg) & (pc) &  [g$_{\mathrm{AB}}$ (erg s$^{-1}$)] & [erg s$^{-1}$ cm$^{-2}$ (erg s$^{-1}$)] \\
\hline \hline
ZTF J0546+3843 & $7.94691(1)$ & 86.6142  & 38.7204 & $3707^{+1631}_{-1258}$ & $19.31\pm{0.0068}$  & $<2.1 \times 10^{-13}$ \\
 & & & & & ($2.55^{+2.7}_{-1.4}\times 10^{33}$) & ($<2.11^{+3.7}_{-1.9}\times 10^{32}$) \\ 
\hline
ZTF J1858--2024 & $8.67990(1)$ & 284.5248 & --20.4135 & $2895^{+2733}_{-1449}$ & $19.37\pm 0.012$ & $<1.1 \times 10^{-13}$ \\
 & & & & & ($2.03^{+5.6}_{-1.5} \times 10^{33}$) & ($<1.10^{+3.1}_{-0.83} \times 10^{32}$)  \\
\hline
ZTF J0425+3858 &  $13.154(1)$ & 66.4592 & 38.9827 & --- & $21.62\pm 0.19$ & $<1.3 \times 10^{-13}$  \\
\end{tabular}
\\
\caption{System parameters for the three sources. Geometric distances derived from Gaia Data Release 3 are quoted from \cite{BailerJones2021}. g-band optical magnitudes are quoted from Gaia DR3 for ZTF J0546+3843 and ZTF J1858--2024, and ZTF for ZTF J0425+3858. Optical luminosities are calculated by integrating a power-law fit to the flux-calibrated LRIS spectra from 3000--10000~\AA. X-ray flux upper limits were obtained via Swift Target-of-Opportunity observations.} \label{tab:objects}
\end{table*}

\subsection{High-speed photometry and phase coherent timing}

The sources presented in this work were discovered by a bulk periodicity search \citep{Burdge2020a} of data from the Zwicky Transient Facility (ZTF; \citealt{Bellm2019,Graham2019,Masci2019,Dekany2020}). We obtained high-speed photometric follow-up observations of our three targets to confirm their orbital periods and determine their timing solutions. We used four high-speed photometers: HiPERCAM, ULTRACAM, CHIMERA, and Lightspeed. HiPERCAM \citep{Dhillon2021} is a frame-transfer, quintuple-beam imager on the 10.4-m Gran Telescopio Canarias (GTC) at Roque de los Muchachos Observatory, which can simultaneously image in u$_s$/g$_s$/r$_s$/i$_s$/z$_s$ bands at rates $>1$ kHz. ULTRACAM \citep{Dhillon2007} is a three-channel high speed photometer mounted on the 3.5-m New Technology Telescope at La Silla Observatory. CHIMERA \citep{Harding2016} is a two-channel photometer on the 200-inch Hale telescope at Palomar Observatory. Lightspeed is a high speed imager under construction for the Magellan telescopes \citep{BurdgeInPrep}. We obtained 0.6 hours of observations using the Lightspeed camera while it was being tested on the 200 inch Hale telescope at Palomar observatory. The camera is Hamamatsu’s qCMOS sensor, which offers deeply sub-electron readout noise, making it possible to obtain higher time resolution observations of J0546 without paying a prohibitive readout noise penalty. We calibrated the absolute timing solution of Lightspeed using a pulse per second signal from a GPS, and have ensured that it is stable to microsecond precision (and we verified this using observations of the Crab Pulsar). The camera’s first exposure is time tagged by a GPS reference card, and we use timestamps from the internal clock to compute the timestamps of each exposure.

All instruments were operated in frame transfer mode to minimize the readout time between exposures. ZTF J0546+3843 and ZTF J1858--2024 also fell within ZTF Deep-Drill fields, i.e. they each have $\sim 100$ photometric measurements within a span of three nights; this higher-than-usual cadence allowed us to use these data as an independent timing epoch without significant decoherence. All HiPERCAM and ULTRACAM data were reduced using the HiPERCAM reduction pipeline\footnote{\href{https://cygnus.astro.warwick.ac.uk/phsaap/hipercam/docs/html/}{cygnus.astro.warwick.ac.uk/phsaap/hipercam/docs/html/}}, while the CHIMERA and Lightspeed data were reduced with a custom aperture photometry pipeline. Table~\ref{tab:observing_log} contains a full observations log listing the dates of observation, instruments used, filters, exposure times, and total exposure lengths. The unique instrument capabilities of rapid exposures ($<10$~s) and no dead-time between exposures were pivotal in determining eclipse mid-times with sufficient precision to test for long-term period changes.

Timing solutions were obtained by first fitting the single epoch with highest signal-to-noise ratio with a five-term Fourier model, which was the minimum needed to fit the overall light curve profile without overfitting the accretion-induced variability. We used Fourier frequencies corresponding to the first five harmonics of the orbital frequency. We then applied this model to each epoch to determine the eclipse times, then fit a timing model using the nested sampling package UltraNest \citep{Buchner2021}. Except in ZTF J0425+3858 (where the data are significantly sparser than the two shorter-period systems), clear deviations from a constant period were apparent, so we expanded quadratically around the initial best-fit period to account for a period derivative term:
\begin{equation}
\phi (t) = f_0 (t-t_0) + \frac{1}{2} \dot{f}_0 (t-t_0)^2 + \ldots~,
\end{equation}
where $\phi(t)$ is the orbital phase at a given time; $t_0$ is some reference epoch; $f_0 = P_0^{-1}$ is the orbital frequency at $t_0$; and $\dot{f}_0 = -\dot{P}_0 P_0^{-2}$ is the frequency derivative, assumed to be constant in time. We defer the measurement of higher-order terms, e.g. $\ddot{f}$, to future work with a longer timing baseline.

\subsection{Spectroscopic follow-up}

We obtained spectroscopic follow-up (Fig.~\ref{fig:spec}) with the Low Resolution Imaging Spectrometer (LRIS; \citealt{Oke1995}) on the 10-m W.~M. Keck I Telescope at Mauna Kea Observatory, which we reduced using the \texttt{lpipe} pipeline version 2020.09 \citep{Perley2019}. Observations were made with the 1$\times$175 arcsec long slitmask, the clear filter, 60 second exposures, and 2$\times$2 CCD binning. For the red arm we used the 300/5000 grism, and for the blue arm we used 600/4000. We obtained spectra of ZTF J0546+3843 on Dec 10, 2020; ZTF J1858--2024 on Apr 20, 2023; and ZTF J0425+3858 on Aug 26, 2022. We computed orbital phases by correcting for the period derivatives, then plotted trailed spectra from the LRIS data by phase-folding the barycenter-corrected, time-resolved spectra (Fig.~\ref{fig:trail}).

In addition to optical spectroscopy, we obtained a far-ultraviolet (FUV) spectrum of the shortest period system in the sample of AM CVns, ZTF J0546+3843, using the Space Telescope Imaging Spectrograph (STIS) aboard the Hubble Space Telescope (HST). We used the G140L grating mode 52$\times$2 aperture. The observations were taken on Oct 7, 2022 (PI: Burdge, proposal ID 16689). We compare our FUV spectrum to that of ES Cet, taken from the International Ultraviolet Explorer (IUE) archive\footnote{\href{https://archive.stsci.edu/iue/}{archive.stsci.edu/iue/}} (Fig.~\ref{fig:UV}). To our knowledge, this is the first published UV spectrum of ES Cet. We fit both UV spectra with a power-law continuum model, plus Lorentzian line profiles to obtain fluxes (or upper limits) for the lines of ionized helium, carbon, silicon, and nitrogen commonly observed in accreting white dwarfs (Table~\ref{tab:uv_lines}).

\subsection{X-ray upper limits with Swift/XRT}

We obtained upper limits on the X-ray flux using the Neil Gehrels Swift Observatory \citep{Gehrels2004} X-ray Telescope (XRT) instrument. Upper limits were computed from the Swift Living X-ray Point Source Catalog\footnote{\href{https://www.swift.ac.uk/LSXPS/}{swift.ac.uk/LSXPS/}} (LSXPS, \citealt{Evans2023}), then converted from counts s$^{-1}$ to energy units by assuming a 1 keV blackbody spectrum (motivated by the X-ray spectrum of the 10.3-minute orbital period AM CVn ES Cet, \citealt{Strohmayer2004b}). The upper-limits are reported in Table~\ref{tab:objects}, and of order $10^{-13}$ erg cm$^{-2}$ s$^{-1}$. 

This is a surprisingly deep non-detection of X-rays. Assuming that matter falls from the inner Lagrange point (L1) to the boundary layer of the accretor, and that half of its gravitational potential energy is dissipated in the accretion disk, the emitted X-ray luminosity ($L_X$) is:
\begin{equation}
L_X \approx \frac{GM_1\dot{M}}{2R_1}\bigg(1-\frac{R_1}{R_{\mathrm{L}1}}\bigg)
\end{equation}
where $M_1$ is the accretor mass, $R_1$ is the donor radius, $R_{\mathrm{L}1}$ is the donor Roche lobe radius, and $\dot{M}$ is the mass accretion rate. Rearranging to yield $\dot{M}$ for an observed X-ray flux gives:
\begin{align}
    \dot{M} \approx 3&\times10^{-11}\;M_\odot\;\mathrm{yr}^{-1} \times \bigg(\frac{F_X}{10^{-13}\;\mathrm{cgs}}\bigg)\times\bigg(\frac{d}{3\;\mathrm{kpc}}\bigg)^2 \notag \\
    &\times \bigg(\frac{R_1}{0.01R_\odot}\bigg) \times \bigg(\frac{M_1}{0.8M_\odot}\bigg)^{-1} \times \bigg(1-\frac{R_1}{R_{\mathrm{L}1}}\bigg)^{-1}
\end{align}

This is 3-4 orders of magnitudes smaller than the $\dot{M}$ typical of ultracompact binaries at such short orbital periods (Fig.~\ref{fig:p_mdot}). We are left to conclude that either 1) the X-ray flux in our systems is significantly obscured, perhaps by the inner disk or foreground ISM absorption, which is particularly severe for small blackbody temperatures $\lesssim 100$ eV; or 2) only a very small fraction of the total accretion luminosity is even released in X-rays, for reasons not understood. A similar magnitude discrepancy was also found in ES Cet \citep{Strohmayer2004b} and most AM CVn systems below 30 min \citep{Begari2023}, meaning this tension is a common feature of the shortest-period disk accretors.

\section{Results} \label{sec:results}

\subsection{Optical photometry \& spectroscopy}
\label{subsec:optical}

\begin{figure*}
    \centering
    \includegraphics[width=\textwidth]{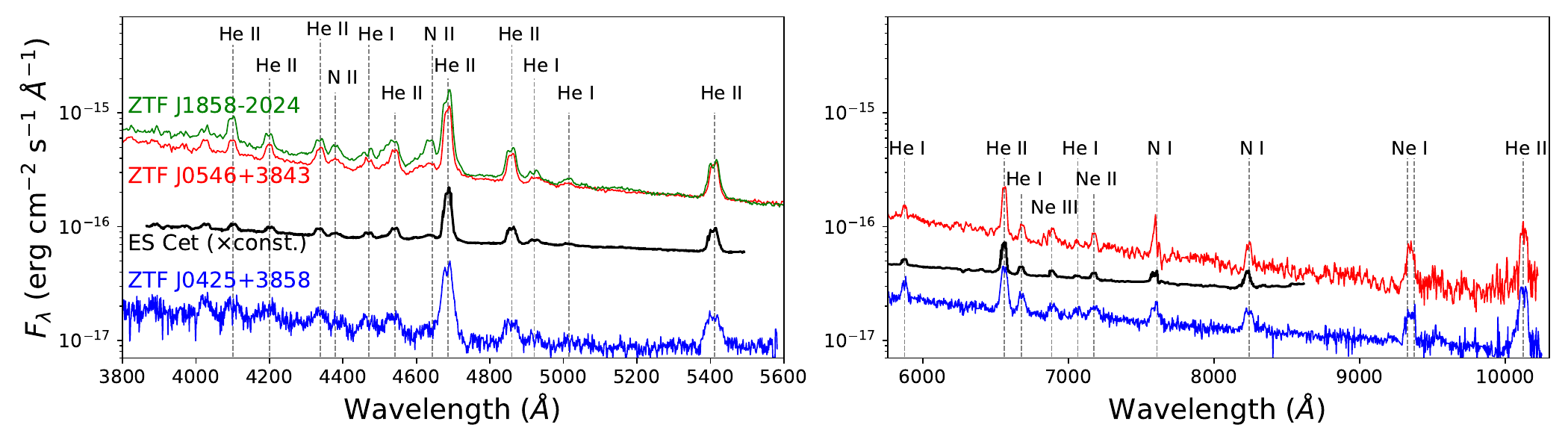}
    \caption{LRIS optical spectra (blue and red arms separated) of new ultracompact binaries, plus archival spectra of ES Cet \citep{Bakowska2021} scaled by an arbitrary normalization constant for ease of comparison. All systems are dominated by double-peaked emission lines of ionized helium, nitrogen, and neon from the accretion disk. The offsets in normalization across the red and blue arms are due to flux-calibration systematic error.}
    \label{fig:spec}
\end{figure*}

\begin{figure*}
    \centering
    \includegraphics[width=\textwidth]{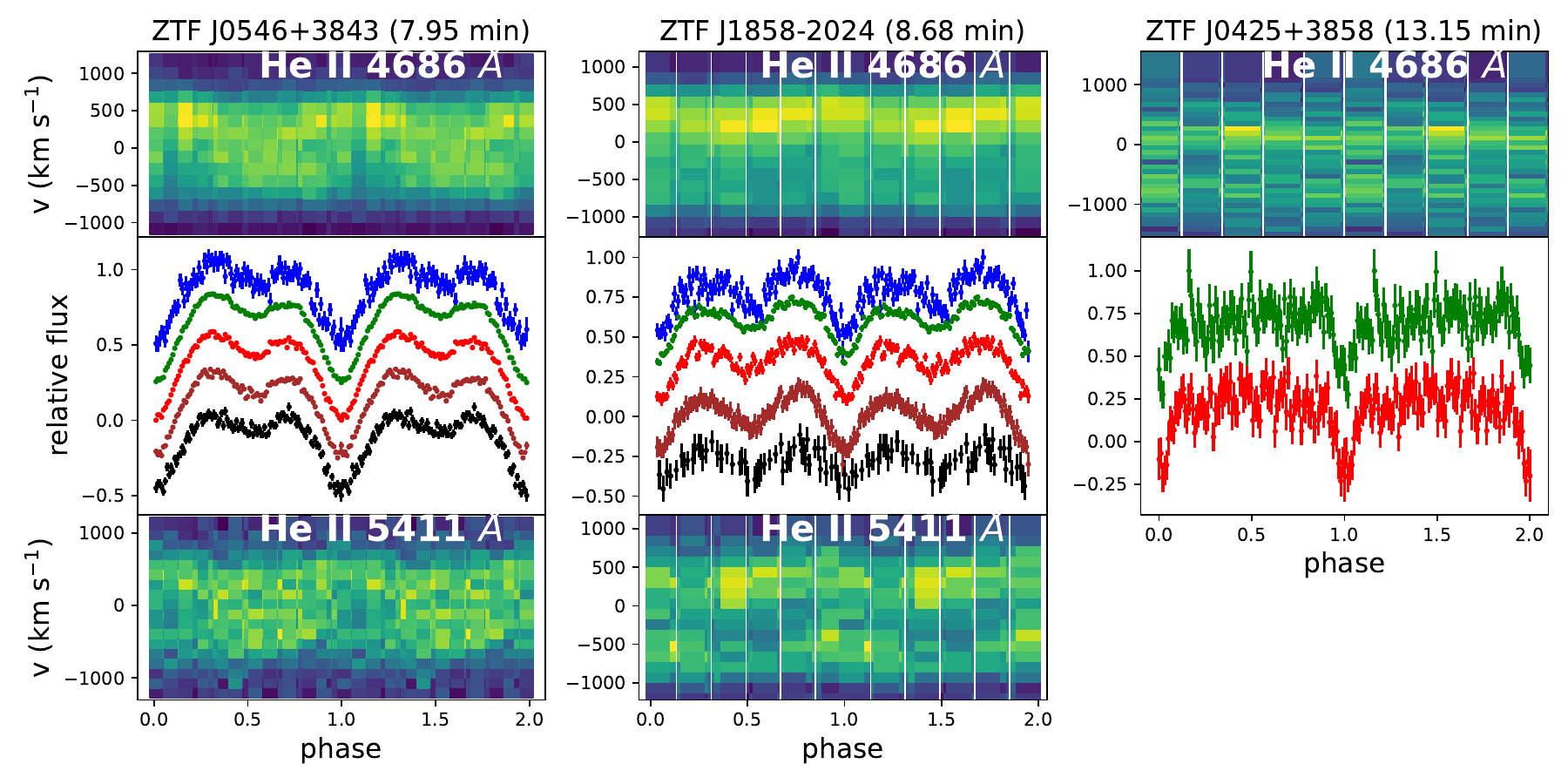}
    \caption{\textbf{Top and bottom panels:} Trailed spectra (continuum-subtracted) of the He\,{\sc ii}~4686 and 5411~\AA\ emission lines that might show S-wave variability at the orbital period. The orbital cycle is plotted twice. All spectra were taken with the LRIS instrument on Keck. \textbf{Middle panels:} phase-folded light curves in different filters. The pictured light curves of ZTF J0546+3843 and ZTF J1858--2024 were taken with HiPERCAM ($u_s$/$g_s$/$r_s$/$i_s$/$z_s$ filters from top to bottom), while ZTF J0425+3858 was observed with CHIMERA ($g$/$r$). Different bands are offset for visual clarity.}
    \label{fig:trail}
\end{figure*}

In Fig.~\ref{fig:spec} we show optical spectra of the three new ultracompact binaries ZTF J0546+3843, ZTF J1858--2024, and ZTF J0425+3858. All three systems are dominated by strong Keplerian profiles of doubly ionized He, N, and Ne, as well as weak He\,{\sc i} emission, indicating a hot, optically thin accretion disk. These lines are all labelled in Fig.~\ref{fig:spec}.

From our photometric follow-up, we find orbital periods of 7.95, 8.68, and 13.15 minutes, which we report alongside the optical X-ray fluxes/upper limits in Table~\ref{tab:objects}. In Fig.~\ref{fig:trail} we show the photometric data phase-folded on the orbital periods in different bands. All systems show significant variability in flux, by 40-60\% through each orbit. While these sources are too faint to have robust parallax measurements with \textit{Gaia} \citep{Gaia2016}, we quote the geometric distance estimates (where available) from \cite{BailerJones2021}. ZTF J0425+3858 was too faint for even a \textit{Gaia} detection, so we have no distance estimate.

Fig.~\ref{fig:trail} also shows the optical spectra in a phase-resolved sequence (trailed spectra), for the strongest emission lines (He\,{\sc ii}~4686 \AA\ and He\,{\sc ii}~5411 \AA). One feature in the trailed spectra is the presence of sinusoidal trails of brightness peaks and troughs offset from the line centroid at zero velocity, known as S-wave signatures (though the feature is less clear in the lower-quality spectrum of ZTF J0425+3858). S-waves are induced by the co-rotation of the disk-impact/bright-spot region near the outer disk edge with the binary orbit \citep{Smak1985}, and they constitute the most robust evidence that the photometric period is indeed the orbital period of these systems.

While it is tempting to proceed by measuring emission-line radial velocities (RVs), the presence of a disk significantly complicates our ability to infer physical properties of the systems from RVs. They would be contaminated by the uncertain dynamics of the accretion disk and bright spot, and it is unclear which emission components contribute most strongly to the observed Doppler shifts. Then, too, it is uncertain which radii within the disk dominate the contribution to the Keplerian line profile; directly translating measured RVs to system parameters of interest is thus unfeasibly complicated (though some limits could be obtained using Doppler tomography, which would require significantly higher-quality spectra than present). Not all hope is lost, as we can still significantly constrain the physical properties of systems by arguments related to their ultracompact orbits (Sec.~\ref{subsec:mass_constraints}), and even further in the two systems with direct measurements of $\dot{P}$ (Sec.~\ref{subsec:pdot_constraints}).

\begin{figure}
    \centering
    \includegraphics[width=\linewidth]{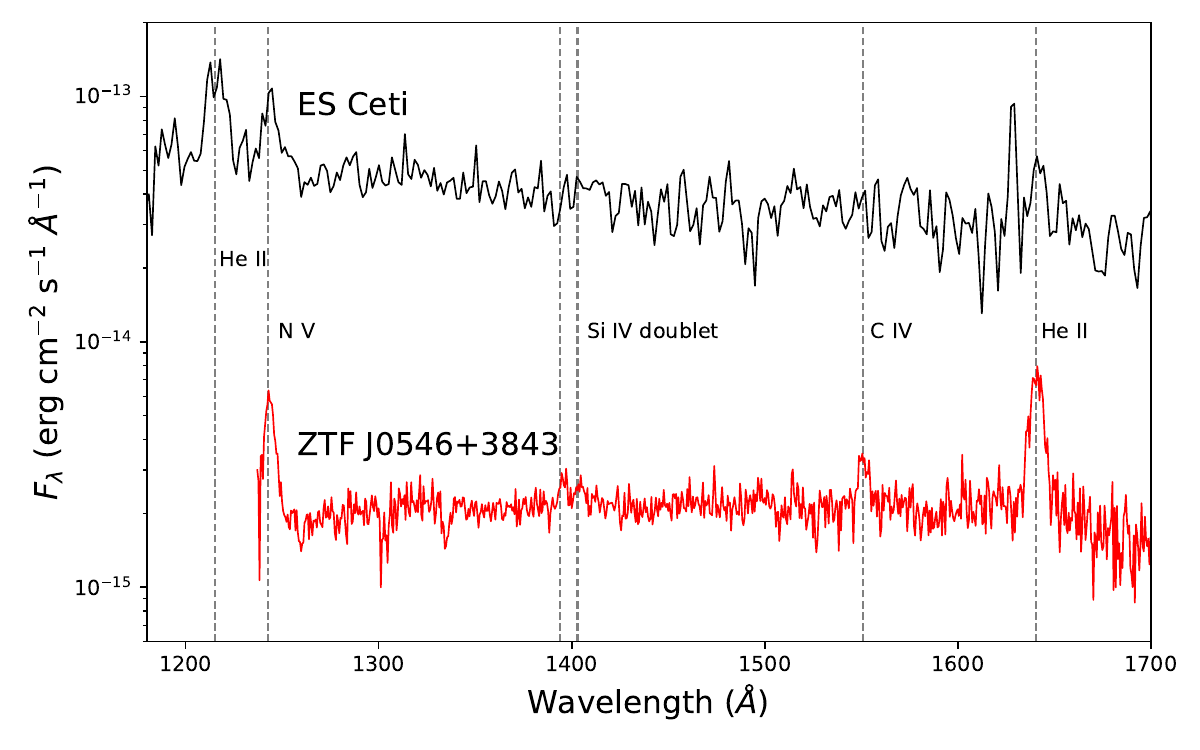}
    \caption{A comparison of the UV spectra of ZTF J0546+3843 (HST STIS) and ES Cet (IUE). Both objects exhibit detectable nitrogen and helium features (though the strength of the nitrogen emission lines relative to helium is much larger in ES Cet). The carbon emission line visible in the spectrum of ZTF J0546+3843 is notably absent in the spectrum of ES Cet.}
    \label{fig:UV}
\end{figure}

\subsection{Ultraviolet spectroscopy}
For ZTF J0546+3843 we also obtained an HST STIS UV spectrum, which is shown in Fig.~\ref{fig:UV} in comparison to the spectrum of ES Cet. The UV spectra of both objects show emission lines of He\,{\sc ii} and N\,{\sc v}. In ZTF J0546+3843, we also detect Si\,{\sc iv} and C\,{\sc iv}, the latter of which is particularly interesting as a direct measure of the degree of CNO processing the donor has undergone (Sec.~\ref{subsec:spectra}). We note that our spectrum loses signal around 1250$\AA$, so we are insensitive to the detection of Ly $\alpha$, which was detected in the inspiraling system HM Cnc \citep{Munday2023}. From the UV lines of N and C, we infer a N\,{\sc v}/C\,{\sc iv} line ratio of 2.3 in ZTF J0546+3843, compared to a lower limit of N\,{\sc v}/C\,{\sc iv}$>5.6$ in ES Cet (we can only quote a lower limit, as C\,{\sc iv} was not detected in ES Cet). It is surprising that the optical spectrum of ZTF J0546+3843 appears identical to that of ES Cet \citep{Bakowska2021}, yet lines uniquely accessible in the UV reveal differing chemical abundances. This emphasizes the importance of far-UV spectroscopy in distinguishing AM CVn evolutionary stages and channels, and the need for future UV spectroscopy missions such as UVEX \citep{Kulkarni2021} and CASTOR \citep{Cote2019}.

\begin{table}[htb]
\centering
Far-ultraviolet emission line fluxes
\begin{tabular}{c|c|c}
\hline \hline
Line & ZTF J0546 (7.9 min) & ES Cet (10.3 min) \\
 & $10^{-15}$ erg s$^{-1}$ cm$^{-2}$ &  $10^{-14}$ erg s$^{-1}$ cm$^{-2}$ \\
\hline
N\,{\sc v} 1242.8\AA & $25.7\pm 0.6$ & $27.6\pm 1.8$  \\
Si\,{\sc iv} 1393.8\AA & $4.18\pm 0.9$ & $<1.8$ \\
Si\,{\sc iv} 1402.8\AA & $6.04\pm 1.3$ & $<4.3$ \\
C\,{\sc iv} 1550.8\AA & $11.4\pm 0.7$ & $<5.0$ \\
He\,{\sc ii} 1640.5\AA & $47.5\pm 0.7$ & $22.3\pm 2.7$ \\
\end{tabular}
\caption{Integrated far-UV emission line fluxes for ZTF J0546+3843 (HST STIS) and ES Cet (IUE) derived from the spectra displayed in Fig.~\ref{fig:UV}.} \label{tab:uv_lines}
\end{table}

\subsection{Long-term timing results}
\label{subsec:timing_results}
\begin{figure*}
    \centering
    \includegraphics[width=\linewidth]{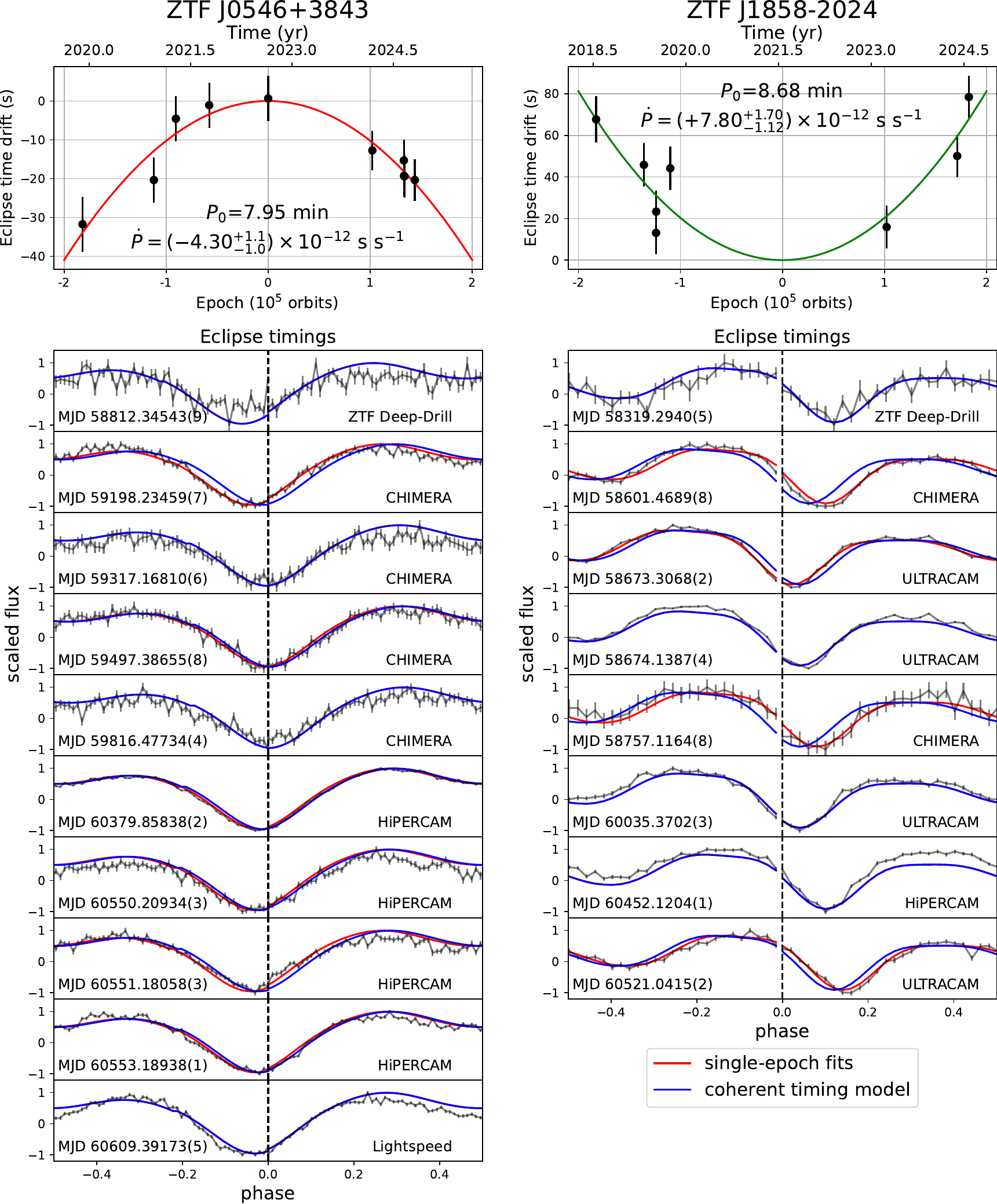}
    \caption{\textbf{Top panels:} $O-C$ diagrams for ZTF J0546+3843 (left) and ZTF J1858--2024 (right). \textbf{Bottom panels:} phase folds with a fixed $P_0$, with eclipse times and instruments noted for each epoch. The eclipse time drifts by $\sim 0.1$-$0.15P$ over the baseline. We overplot the predicted waveforms from our coherent timing models (blue), as well as the single-epoch fits used to construct the $O-C$ diagrams. The 5-10 s discrepancies are likely dominated by timing jitter due to accretion (see Sec.~\ref{subsec:timing_results}).}
    \label{fig:timing}
\end{figure*}

In Fig.~\ref{fig:timing} we show the $O-C$ diagrams corresponding to our best-fit timing solutions for the 7.9- and 8.7-minute systems. We infer $t_0=59816.4773(1)$ MJD, $P_0=7.94691(5)$ min, and $\dot{P}=(-4.30^{+1.1}_{-1.0})\times 10^{-12}$ s s$^{-1}$ for ZTF J0546+3843; and $t_0=59420.0012(1)$ MJD, $P_0=8.67990(3)$ min, and $\dot{P}=(+7.80^{+1.70}_{-1.12})\times 10^{-12}$ s s$^{-1}$ for ZTF J1858--2024. It is noteworthy that the period derivatives have opposite signs, despite similar optical spectra and orbital periods. The inspiral seen in ZTF J0546+3843 is driven by the familiar effects of GR: gravitational waves remove angular momentum from the system, shrinking the binary orbit. The physical driver of the expanding orbit of ZTF J1858--2024 is less obvious. This is an effect seen directly in only one other ultracompact binary so far \citep{deMiguel2018}, though it is expected to happen eventually in every AM CVn as a consequence of angular momentum conservation during mass transfer from a low-mass donor to a high-mass accretor (see Sec.~\ref{subsec:pdot_constraints} for details). Thus, in the standard evolutionary picture, initially the outermost high-entropy layers are stripped, tending to produce contraction and negative $\dot{P}$. This is gradually dominated by the expansion of the underlying layers, causing the donor to expand and the system to outspiral \citep{Deloye2007}.

In Fig.~\ref{fig:timing}, we see that the coherent timing models (blue) are not always a perfect match to the epoch-by-epoch phase files (black datapoints and red models). This is likely due to the optical emission being dominated by the accretion disk: as a result, short-term changes in the accretion flow (flicker noise) can result in the center of light shifting by $\sim 5-10$\% of the orbital period \citep{deMiguel2018,Munday2023}. However, the magnitude of this flickering effect is extremely unlikely to explain the long-term trend of 40-80 s drift in the eclipse timings caused by period evolution (Fig.~\ref{fig:timing}).

\section{Discussion} \label{sec:discussion}

\subsection{Spectroscopic properties and evolutionary channels}
\label{subsec:spectra}

As reviewed in Sec.~1, the chemical abundances of AM CVns carry imprints of their formation channels, in particular the differences in C/O, N/O, and N/C ratios due to donors which have undergone varying levels of He and CNO fusion \citep{Nelemans2010}. To this end, far-ultraviolet (FUV) spectroscopy is particularly rewarding, as the highest density of diagnostic metal-line transitions resides in the FUV. Following the approach of \cite{Marsh1995} and \cite{Gansicke2003}, we use the ratio of N\,{\sc v}/C\,{\sc iv} as a diagnostic of CNO processing rather than directly fitting for the N/C or N/He ratios, which would require full LTE modeling which is beyond the scope of this work. 

We measured a line ratio N\,{\sc v}/C\,{\sc iv}=2.3 in ZTF J0546+3843 and N\,{\sc v}/C\,{\sc iv}$>5.6$ in ES Cet (Fig.~\ref{fig:UV}, Table~\ref{tab:uv_lines}). For comparison, typical long-period CVs show N\,{\sc v}/C\,{\sc iv}$\approx$0.3. FUV spectroscopy of ultracompact binaries is sparse, but as an example the 46-minute system GP Com shows a significantly higher N\,{\sc v}/C\,{\sc iv}=10 \citep{Marsh1995}. The donor in ZTF J0546+3843 has thus undergone significantly more CNO processing than typical CVs with $P \gtrsim 80$ min, but slightly less than the longer-period GP Com and ES Cet. 

Modest variations in the equilibrium CNO abundances from fusion are expected for different main sequence progenitors of the WD donors in these systems. Moreover, gravitational settling will tend to stratify the CNO-burning products within the WD core itself, meaning donors that have been transferring mass for longer (and are therefore more stripped) can show changes in N/C ratio over their lifetimes. The lower N/C measurement in ZTF J0546+3843 therefore does not necessarily mean it descended from the He star channel. Most problematically, such systems are expected to reach period minima around 9-11~min \citep{Yungelson2008}, slightly longer than the 7.95~min orbital period observed in the system. We speculate the ZTF J0546+3843 descends from a He WD donor whose main sequence progenitor was less massive than that of ES Cet or GP Com, and is also less stripped as it is earlier in its mass-transferring lifetime (inspiral rather than outspiral phase). 

Our ability to make similar inferences for ZTF J1858--2024 and ZTF J0425+3858 is limited by our lack of UV spectra of those sources. Our results tentatively support the idea that AM CVns showing positive $\dot{P}$ preferentially exhibit greater N/C abundances than those with negative $\dot{P}$, in part due to gradual stripping of the donor via mass transfer. This speculation can be confirmed by obtaining high-resolution UV spectroscopy of ZTF J1858--2024 which is known to be outspiraling, thus must be more evolved than ZTF J0546+3843.

The extreme similarity between the optical spectra of the inspiraling system ZTF J0546+3843, and the outspiraling systems ZTF J1858--2024 and ES Cet, is surprising, especially given the non-detection of H in ZTF J0546+3843. Evolutionary models of double white dwarfs typically assume the initial inspiral phase after mass-transfer contact consists primarily of stripping the thin, thermally inflated H/He envelope remaining on the He WD surface from the progenitor main-sequence star \citep{DAntona2006,Deloye2007,Kaplan2012,Wong2021,Chen2022}. This layer is non-degenerate, and it contracts upon mass loss, which allows the donor radius to shrink rapidly as the system undergoes orbital decay. Eventually the H/He shell is depleted, leaving only the degenerate core which expands upon mass-loss resulting in a net positive $\dot{P}$.

ZTF J0546+3843 complicates this picture, showing both a negative $\dot{P}$ and no H. One clue towards an explanation comes from considering the double WD evolutionary models of \cite{Kaplan2012}: their Eq. 10 finds that $\dot{P}_{\mathrm{in}}$ for inspiraling systems prior to period minimum and turnaround should be approximately $-1.7\times 10^{-11}$ s s$^{-1}$ (for $P=$7.95 min, and assuming typical donor/accretor masses of 0.15/0.8 $M_\odot$). This is $>35\sigma$ larger in magnitude than our measured $\dot{P}=-4.30^{+1.1}_{-1.0}\times 10^{-12}$ s s$^{-1}$, leading us to speculate that ZTF J0546+3843 is well past its loss of a H envelope and nearly turning around to a positive $\dot{P}$. In this scenario, most of the H shell must have already been stripped, resulting in an extremely low surface H abundance resulting in the nondetection.

\subsection{Mass constraints based on period}
\label{subsec:mass_constraints}

As discussed in Section \ref{subsec:spectra}, it is difficult to make informed conclusions about the donor or accretor masses because the optical emission is entirely dominated by the accretion disks in all of these systems. However, we can still make some inferences to place strict lower limits, and make some estimates, on the possible accretor and donor masses.

\textbf{Accretor mass bound:} Observing the presence of an accretion disk in a system (via double-peaked line profiles) means that a system is not undergoing direct-impact accretion. In other words, the radius of the accretor must be small enough that the ballistic trajectory of the accretion stream through the first Lagrange point (L1) does not intersect the accretor outer radius. This upper bound on the radius translates to a lower bound on the mass, which is given by assuming a zero-temperature electron-degenerate equation of state (an arbitrarily more massive white dwarf could attain the same radius by having a nonzero temperature.) We use Eq.~6 of \cite{Nelemans2001} to estimate this maximum accretor radius in units of the orbital separation:
\begin{align}
    \frac{R_1}{a} &\lesssim 0.04948 - 0.03815\,\log(q) \nonumber \\
    &\quad + 0.04752\,\log^2(q) - 0.006973\,\log^3(q)~,
\end{align}
where $q\equiv M_2/M_1$ is the mass ratio for donor and accretor masses of $M_2$ and $M_1$ respectively. We then cite the zero-temperature white dwarf mass-radius relation quoted by \cite{Verbunt1988}:
\begin{align}
    \frac{R}{R_\odot} &= 0.0114\bigg[\bigg(\frac{M}{M_\textrm{Ch}}\bigg)^{-2/3} - \bigg(\frac{M}{M_\textrm{Ch}}\bigg)^{2/3}\bigg]^{1/2} \nonumber \\
    &\quad\quad\times \bigg[1+3.5\bigg(\frac{M}{\mathrm{M}_\textrm{p}}\bigg)^{-2/3} + \bigg(\frac{M}{\mathrm{M}_\textrm{p}}\bigg)^{-1}\bigg]^{-2/3}~, 
    \label{eq:mass_radius_relation}
\end{align}
where $M$ is the accretor mass, $M_\mathrm{Ch}$ denotes the Chandrasekhar limit for the mass of a white dwarf, which is 1.44\,$M_\odot$, and $\mathrm{M}_\mathrm{p}$ is a constant with a value of 0.00057\,$\mathrm{M}_\odot$, to convert this to an accretor mass constraint (see Sec. 4 of \citealt{Verbunt1988}). We emphasize that although white dwarfs realized in nature deviate significantly from the zero-temperature relation (e.g. Fig. 11 of \citealt{Green2018a}, Fig. 12 of \citealt{vanRoestel2022}), our purpose in quoting it here is for an absolute 
physically allowed lower bound. Attempting to fit a lower-mass object into this orbit is guaranteed to result in direct-impact, rather than disk, accretion.

\textbf{Donor mass bound:} 
For the donor mass, we can use the relation of \cite{Eggleton1983}, in a simplified form valid for the relevant mass-ratio range of CVs ($0.01\lesssim q \lesssim1$):
\begin{equation}
    P_{\mathrm{RLO}} \approx 6.192 \; \mathrm{min} \times \bigg(\frac{\rho}{10^4 \mathrm{\;g\;cm}^{-3}}\bigg)^{-1/2}~,
    \label{eq:eggleton}
\end{equation}
where $P_{\mathrm{RLO}}$ is the orbital period at which an object of mean density $\rho$ overflows its Roche lobe. Again assuming the zero-temperature  equation of state quoted above gives the minimum possible mass to attain this density.

\begin{figure}
    \centering
    \includegraphics[width=\linewidth]{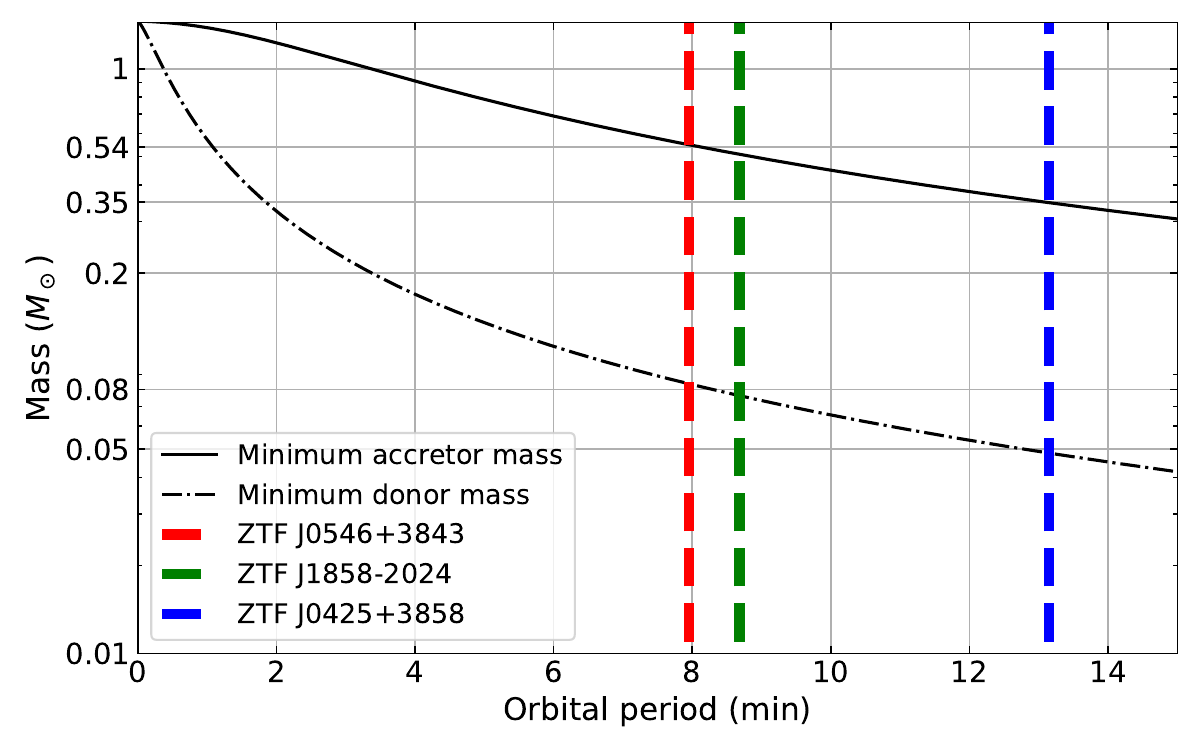}
    \caption{Minimum mass bounds for disk-accreting systems as a function of orbital period, computed by the argument outlined in Section~\ref{subsec:mass_constraints}. For the relevant orbital periods of 7.95, 8.68, and 13.15 min, we obtain minimum donor/accretor masses of 0.084/0.54, 0.076/0.51, and 0.048/0.35\,$M_\odot$, respectively.}
    \label{fig:mass_constraints}
\end{figure}

We applied these constraints to the orbital periods of our systems. The results are illustrated in Fig.~\ref{fig:mass_constraints}. These yield, for the orbital periods of 7.95, 8.68, and 13.15 min, firm lower limits on the minimum donor/accretor masses of 0.084/0.54, 0.076/0.51, and 0.048/0.35\, $M_\odot$, respectively.

\subsection{Further constraints based on $\dot{P}$} \label{subsec:pdot_constraints}

The direct measurement of the orbital period first derivative, $\dot{P}$ (Fig.~\ref{fig:timing}), is a valuable tool. As an instructive example of how it can significantly constrain the properties of an accreting binary system where angular momentum loss is driven by the well-understood properties of gravitational radiation, we outline the following arguments for ZTF J0546+3843 and ZTF J1858--2024.

Expressing the orbital period of the system in terms of its angular momentum ($J$) and component masses:

\begin{equation}
    P = J^3 (M_1+M_2) (M_1M_2)^{-3} \bigg(\frac{2\pi}{G^2}\bigg)~,
\end{equation}

\noindent
Taking the logarithmic derivative, and assuming approximately steady-state conservative mass transfer ($\dot{M_1}=-\dot{M_2} \equiv
 \dot{M}$, where $\dot{M}$ is strictly positive) yields:

\begin{equation}
    \frac{\dot{P}}{P} = 3\bigg[\frac{\dot{J}}{J} + \frac{\dot{M}}{M_{\mathrm{2}}}(1-q)\bigg]~.
    \label{eq:pdotp_jdotj_mdotm}
\end{equation}

\begin{figure}
    \centering
    \includegraphics[width=\linewidth]{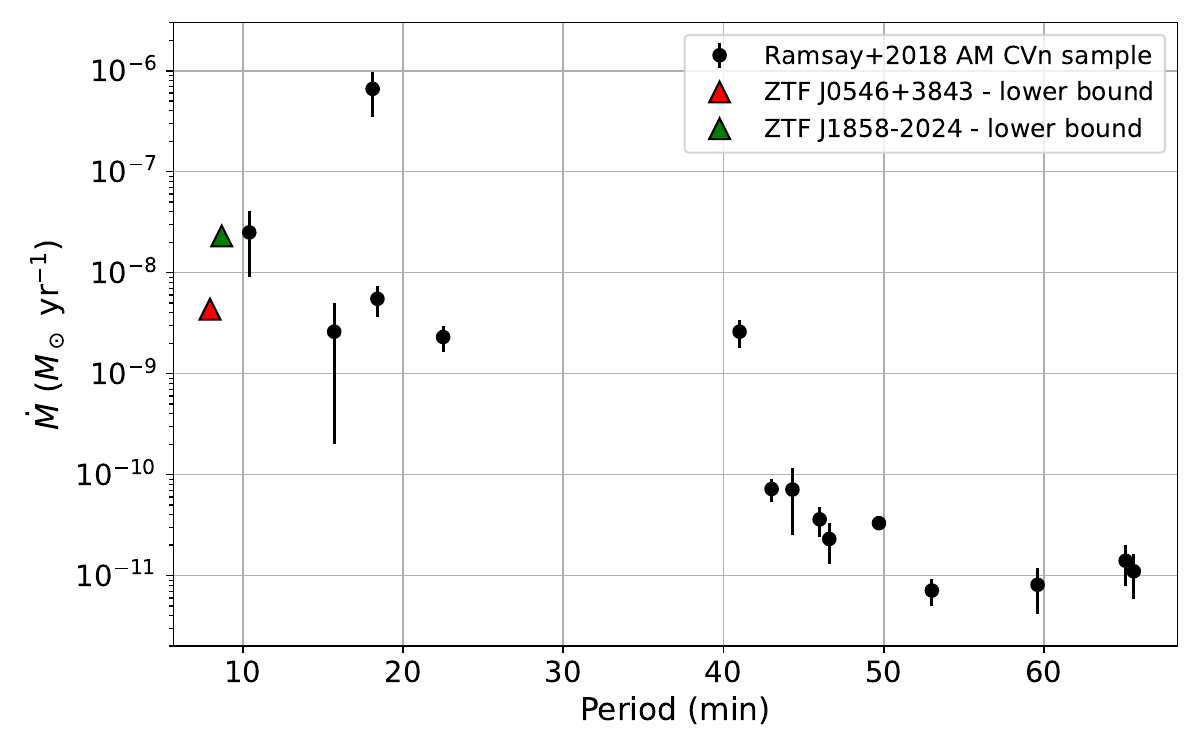}
    \caption{Orbital period vs $\dot{M}$ for AM CVns. \cite{Ramsay2018} obtained their values by fitting UV-optical-IR spectral energy distributions. The lower bounds for ZTF J1858--2024 and ZTF J0546+3843 were obtained by using the lowest allowable masses (via the argument in Sec.~\ref{subsec:mass_constraints}) and Eq.~\ref{eq:mdot}.}
    \label{fig:p_mdot}
\end{figure}

The rate of change in orbital angular momentum is driven by three effects: gravitational radiation, angular momentum loss in the mass-transfer stream, and tidal coupling of the accretor/disk/donor, which scales with the difference between accretor spin angular momentum and binary orbital angular momentum \citep{Marsh2004}. In the presence of a radially extended accretion disk (as opposed to a direct-impact stream), it is generally assumed that tidal coupling is efficient, so that the accretor spin is kept close to the orbital frequency and the angular momentum lost in the accretion stream is quickly returned to the orbit \citep{Verbunt1988,Frank2002}. In this case, the angular momentum evolution of the system is set entirely by gravitational wave emission ($\dot{J} = \dot{J}_{\mathrm{GR}}$). We quote the standard expression for $\dot{J}_{\mathrm{GR}}$ in an inspiraling binary system \citep{Peters1964}:
\begin{align}
    \bigg(\frac{\dot{J}}{J}\bigg)_{\mathrm{GR}} &= -\frac{32}{5} \frac{G^{5/3}M_1 M_2}{c^5 (M_1+M_2)^{1/3}} \bigg(\frac{2\pi}{P}\bigg)^{8/3} \\
    &= -\frac{32}{5}\bigg(\frac{G\mathcal{M}_c}
    {c^3}\bigg)^{5/3}\bigg(\frac{2\pi}{P}\bigg)^{8/3}~.
\end{align}
Thus, the inferred mass transfer rate for a disk-accreting system is:
\begin{equation}
    |\dot{M}| = \frac{M_2}{1-q} \bigg[\frac{\dot{P}}{3P} - \bigg(\frac{\dot{J}}{J}\bigg)_{\mathrm{GR}}\bigg]~.
    \label{eq:mdot}
\end{equation}
This is already a useful result: using the strict lower bounds on $M_1$, $M_2$ (Fig.~\ref{fig:mass_constraints}) and the measured values of $P$/$\dot{P}$ (Fig.~\ref{fig:timing}) yields approximate lower bounds of $\dot{M} \gtrsim 4.3\times 10^{-9}$ $M_\odot$ yr$^{-1}$/$2.5\times 10^{-8}$ $M_\odot$ yr$^{-1}$ for ZTF J0546+3843/ZTF J1858--2024 respectively. These are reasonable values compared to extrapolation from systems at similar orbital periods, which we illustrate in Fig.~\ref{fig:p_mdot}.

We can further eliminate the $\dot{M}$-dependence of Eq.~\ref{eq:pdotp_jdotj_mdotm} by considering the response of the donor star to mass loss. We begin by assuming steady-state contact of the donor with its Roche lobe, meaning $R_2=R_L$ at all times (where $R_L$ denotes the Roche lobe radius). For simplicity, we use the \cite{Paczynski1971} result for $R_L$,
\begin{equation}
    \frac{R_L}{a} = 0.46\bigg(\frac{M_2}{M_1+M_2}\bigg)^{1/3}~,
    \label{eq:roche}
\end{equation}
where $a$ is the binary semimajor axis. We take the logarithmic derivative of Eq.~\ref{eq:roche}, replace $R_L=R_2$, express $a$ in terms of $P$, and assume conservative mass transfer ($\dot{M}_\mathrm{1}+\dot{M}_2=0$) to obtain:
\begin{equation}
    \frac{2\dot{P}}{3P} - \frac{|\dot{M}|}{3M_2} = \bigg(\frac{\dot{R_2}}{R_2}\bigg) = \bigg(\frac{\dot{R_2}}{R_2}\bigg)_{\mathrm{ad}} + \bigg(\frac{\dot{R_2}}{R_2}\bigg)_{\mathrm{evol}}~.
    \label{eq:rdot_split}
\end{equation}

On the right hand, we split $(\dot{R_2}/R_2)$ into a portion describing the adiabatic (short-term) response of the donor to mass loss, and a portion describing the long-term radius evolution of the donor star driven by thermal contraction and/or nuclear burning. We define the adiabatic response in terms of the index $\xi_{\mathrm{ad}}$:
\begin{equation}
    \xi_{\mathrm{ad}} \equiv \frac{d(\ln R_2)}{d(\ln M_2)}\bigg|_{\rm ad} =  \frac{M_2}{R_2}\frac{dR_2}{dM_2}\bigg|_{\rm ad} = -\bigg(\frac{\dot{M}}{M_2}\bigg)^{-1}\bigg(\frac{\dot{R}_2}{R_2}\bigg)_{\rm ad}~,
    \label{eq:adiabatic_index}
\end{equation}
which for low-mass white dwarfs is very near $-1/3$. We also cast the longer-term evolution in terms of a dimensionless parameter:
\begin{equation}
    \chi_{\mathrm{evol}} \equiv \frac{\tau_{\mathrm{GR}}}{\tau_{\mathrm{evol}}} = \frac{(\dot{R_2}/{R_2})_{\mathrm{evol}}}{(\dot{J}/{J})_{\mathrm{GR}}}
    \label{eq:chi}~,
\end{equation}
which compares the donor thermal/nuclear contraction rate to the gravitational wave-driven orbital evolution rate. $\chi_{\mathrm{evol}} \ll 1$ corresponds to systems in which the GR decay is much faster than the donor evolution due to thermal contraction/nuclear burning (e.g., LIGO mergers), and $\chi_{\mathrm{evol}} \gg 1$ corresponds to systems where the donor contraction quickly brings the system out of Roche lobe contact.

\begin{figure}
    \centering
    \includegraphics[width=\linewidth]{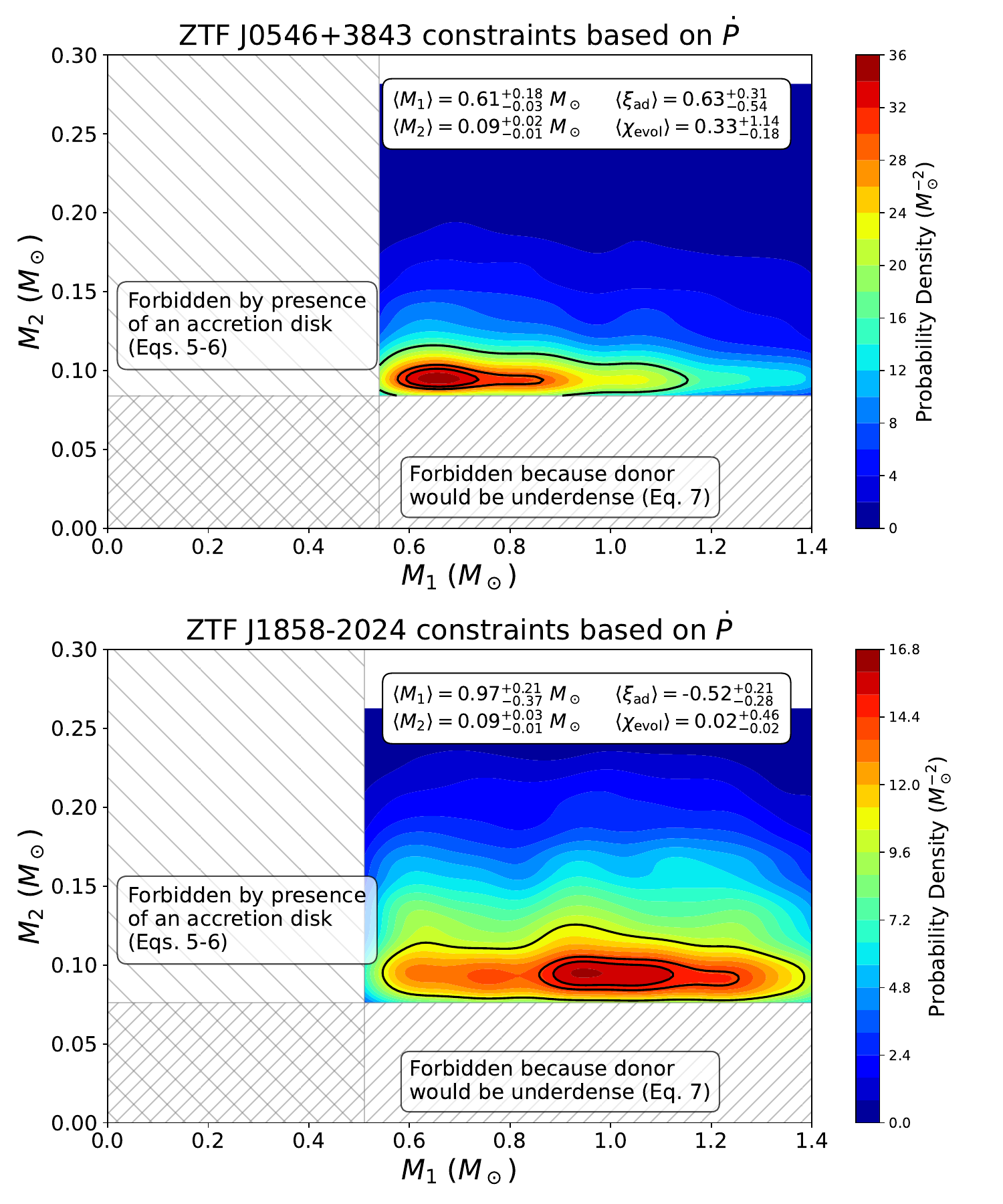}
    \caption{Likelihood contours of $M_1$ and $M_2$ for our Monte Carlo runs based on the argument outlined in Section~\ref{subsec:pdot_constraints}. Solid black lines correspond to 1/2/3$\sigma$ contours. These imply chirp masses close to $0.3\,M_\odot$ in both systems. As expected from the inward vs. outward $\dot{P}$, the donor of ZTF J0546+3843 is significantly less degenerate ($\xi_{\mathrm{ad}}$ larger) than that of ZTF J1858--2024.}
    \label{fig:pdot_constraints}
\end{figure}

Plugging Eqs.~\ref{eq:adiabatic_index} and \ref{eq:chi} into Eq.~\ref{eq:rdot_split}, subtracting from Eq.~\ref{eq:pdotp_jdotj_mdotm} to eliminate the $\dot{M}/M_2$-dependence, and writing the final result in terms of $\xi_{\mathrm{ad}}$ and $\chi_{\mathrm{evol}}$ gives:
\begin{equation}
    \bigg(\frac{\dot{P}}{P}\bigg)_{\mathrm{eq}} = 3\bigg(\frac{\dot{J}}{J}\bigg)_{\mathrm{GW}}\bigg[\frac{\xi_{\mathrm{ad}} + \chi_{\mathrm{evol}}(1-q) - 1/3}{\xi_{\mathrm{ad}} + 5/3 - 2q} \bigg]~.
    \label{eq:pdotp_jdotj}
\end{equation}
This result expresses the observable quantity $\dot{P}/P$ purely in terms of the system parameters $\xi_{\mathrm{ad}}$, $\chi_{\mathrm{evol}}$, $M_1$, and $M_2$. For mass ratios $0<q\lesssim 2/3$ relevant for most AM CVns, the term in brackets typically spans the range of -1 to 1, with a heavy skew towards positive values (orbital decay).

Further progressing beyond this point would involve assuming values for $\xi_{\mathrm{ad}}$ and/or $\chi_{\mathrm{evol}}$. This is a difficult problem to solve exactly, given our uncertainty on the nature of the donors in these systems (white dwarf vs. He star), as well as general uncertainty in the thermal structure and adiabatic response of degenerate/semi-degenerate matter to rapid mass loss. However, we can loosely constrain the systems by exploring a large range of feasible parameters via a Monte Carlo search. We drew from a random grid of $10^7$ combinations from:
\begin{align*}
    M_1 \sim\mathcal{U}(M_{1,\mathrm{min}},M_{\mathrm{Ch}}) \;&;\; M_2\sim\mathcal{U}(M_{2,\mathrm{min}},M_1) \;;\\ 
    \xi_{\mathrm{ad}}\sim\mathcal{U}(-1, 3) \;&;\; \chi_{\mathrm{evol}}\sim\mathcal{U}(0,2) \;;
\end{align*}
where $M_{1,\mathrm{min}}$, $M_{2,\mathrm{min}}$ are taken from Fig.~\ref{fig:mass_constraints} and $\mathcal{U}$ is the uniform distribution. We retained acceptable parameter combinations if they produced a $\dot{P}$ within $1\sigma$ of the measured values. We also required solutions to fill their Roche lobe, i.e. $R_2 = R_L$. Here, rather than using the zero-temperature mass-radius relation (Eq.~\ref{eq:mass_radius_relation}), we allowed thermal bloating of the donor star by up to 50\% (following the finite-entropy evolutionary models of \citealt{Wong2021}, which found that He WD donors in AM CVn can have this degree of bloating compared to the cold WD equation-of-state close to period minimum). After retaining the acceptable solutions, we used a gaussian kernel density estimator to determine the most probable parameter combinations reproducing the observed system properties; the resulting contours for $M_1$ and $M_2$ are shown in Fig.~\ref{fig:pdot_constraints}.

We obtain mass estimates of $M_1=0.61^{+0.18}_{-0.03}\;M_\odot$ and $M_2=0.09^{+0.02}_{-0.01}\;M_\odot$ for ZTF J0546+3843, and $M_1=0.97^{+0.21}_{-0.37}\;M_\odot$, $M_2=0.09^{+0.03}_{-0.01}\;M_\odot$, for ZTF J1858--2042 (see Fig.~\ref{fig:mass_constraints} for more detailed system parameters). As expected qualitatively, the donor in ZTF J0546+3843 must be semi-degenerate ($\xi_{\mathrm{ad}}>0$), likely a white dwarf with a small inflated envelope, for the orbit to decay \citep{Wong2021}. On the other hand, the donor in ZTF J1858--2024 is consistent with the equilibrium $\xi_{\mathrm{ad}}=-1/3$ for a cold WD.

\begin{figure}
    \centering
    \includegraphics[width=\linewidth]{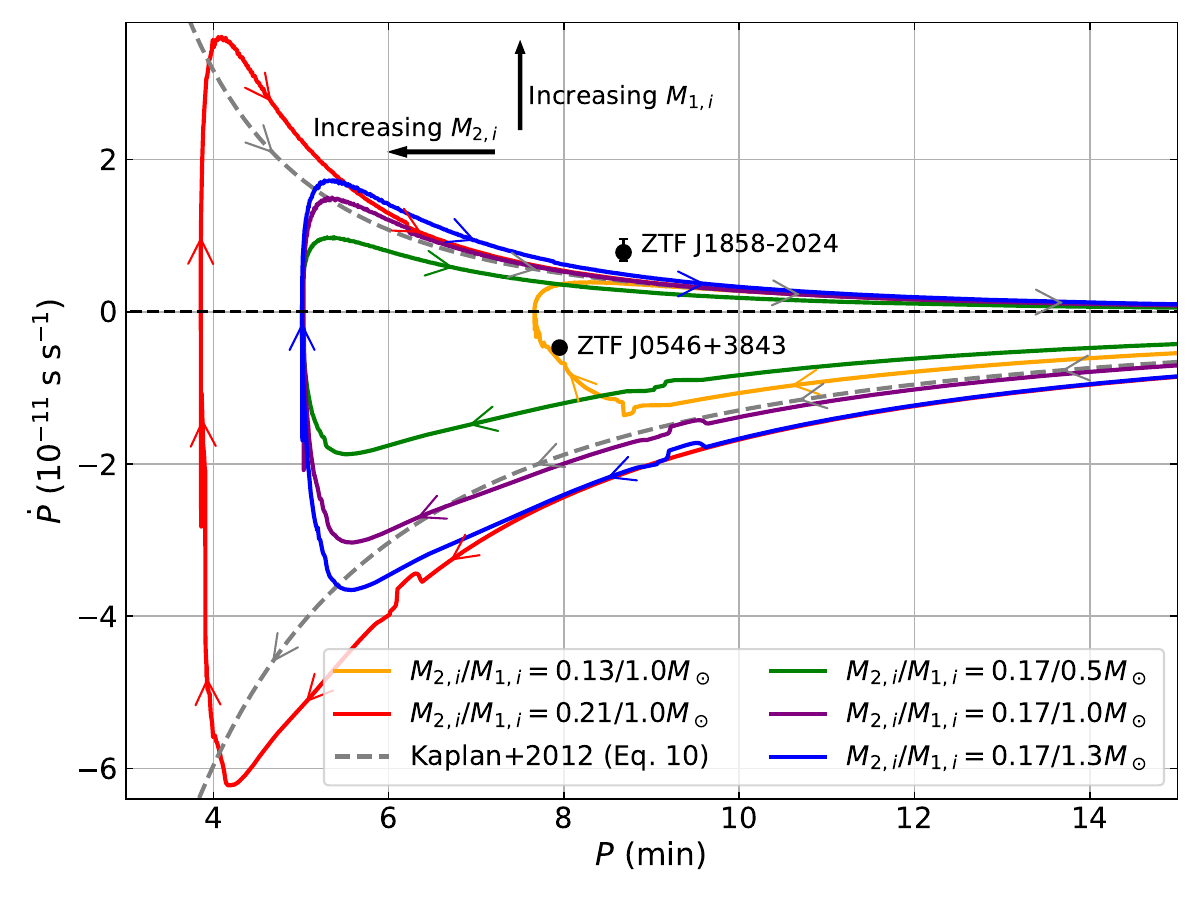}
    \caption{Evolutionary tracks (from \textit{MESA} binary modeling) of double WDs with different binary parameters. The observed $P$ and $\dot{P}$ in ZTF J0546+3843 are compatible with a donor of $M_2\lesssim 0.13 M_\odot$ close to period minimum. The large observed $\dot{P}$ in ZTF J1858+2024 is more difficult to reconcile. In all cases, the donors initially have a hydrogen envelope mass of $7\times 10^{-3}\;M_\odot$.}
    \label{fig:mesa_tracks}
\end{figure}

To further verify our estimates, we ran a small grid of five binary stellar evolution models using \textit{MESA} \citep[version 12115,][]{Paxton2011,Paxton2013,Paxton2015,Paxton2018,Paxton2019}, and also compared to literature results  \citep{Kaplan2012,Chen2022}. In our calculations, the accretor is considered as a point mass. The initial He WD models and assumptions about the mass retention efficiency are adopted from \cite{Chen2022} (see their Sec. 2).
For our models, we assumed initial binary parameters of $M_{1,i}=1.3/1.0/1.0/1.0/0.50\;M_\odot$, $M_{2,i}=0.17/0.21/0.17/0.13/0.17\;M_\odot$, $P_{\rm orb,i}=0.05\;$days, and an initial H envelope mass of $7\times 10^{-3}\;M_\odot$ (where the subscript $i$ indicates the initial value of a parameter; see Fig.~\ref{fig:mesa_tracks} for details). The relatively large initial envelope mass was motivated by the recent direct detection of H in HM Cnc \citep{Munday2023}, suggesting that a significant residual envelope may survive all the way down to $<10$~min periods. We have not explicitly verified that the entire envelope is lost by the periods measured for our systems, as we did not place upper bounds on the H fraction from our optical spectra. The purpose of our model grid is to illustrate the magnitude and direction of changing these parameters: increasing the initial donor mass results in a shorter minimum period, while increasing the initial accretor mass results in a larger overall $|\dot{P}|$ (Fig.~\ref{fig:mesa_tracks}). We note that the discontinuities around 6 and 9 min in these tracks are due to the onset of mass transfer and stripping of the H envelopes. The angular momentum sink of direct-impact accretion is not included in these models, as they are intended to serve as a qualitative illustration of the effects of varying $M_1/M_2$ on the overall trajectories.

The small $\dot{P}$ in ZTF J0546+3843 is constraining: we infer the system is likely close to its period minimum, as it should otherwise have $\dot{P}\lesssim -1.7\times 10^{-11}$ s s$^{-1}$ (our Fig.~\ref{fig:mesa_tracks}; \citealt{Kaplan2012} Fig. 6). Double WD systems which reach period minima at $P \approx 7.9$ minutes require the lowest initial donor masses of around $0.12-0.13 M_\odot$ ($M_i$ cannot be much lower than this, as it is set by the He core mass at the end of the common envelope phase). Our binary evolution models find that the donor mass remains relatively constant for the entire ingoing track, and only begins to change significantly around period minimum (due to the sharp $\dot{M}$ increase); so indeed, our inferred $M_2= 0.09^{+0.02}_{-0.01} M_\odot$ seems compatible with such a low initial mass followed by mass-loss.

On the other hand, our results, and those of \cite{Kaplan2012}, have difficulty reproducing the large observed $\dot{P}$ in ZTF J1858--2024 (except for the smallest values of $M_2$ and the largest values of $M_1$). Our Eq.~\ref{eq:pdotp_jdotj} accommodates this by pushing towards unphysically small values of $\xi_{\mathrm{ad}}<-1/3$, but in reality, we are uncertain exactly how the system is evolving so rapidly. It is possible that finite-entropy donors can accelerate the orbital evolution (e.g. \citealt{Wong2021}), though whether this is a sufficiently large effect to account for the observed discrepancy is unclear. Reproducing such a large $\dot{P}$ via binary evolutionary models is an interesting direction for future study, though observations of $\dot{P}$ may only be expected to agree with models over longer timescales due to the finite time required for the Roche lobe and donor star radius to reach equilibrium.

\subsection{Accreting Ultracompact Binaries as Multi-Messenger GW sources}
\label{subsec:multimessenger}

\begin{figure}
    \centering
    \includegraphics[width=\linewidth]{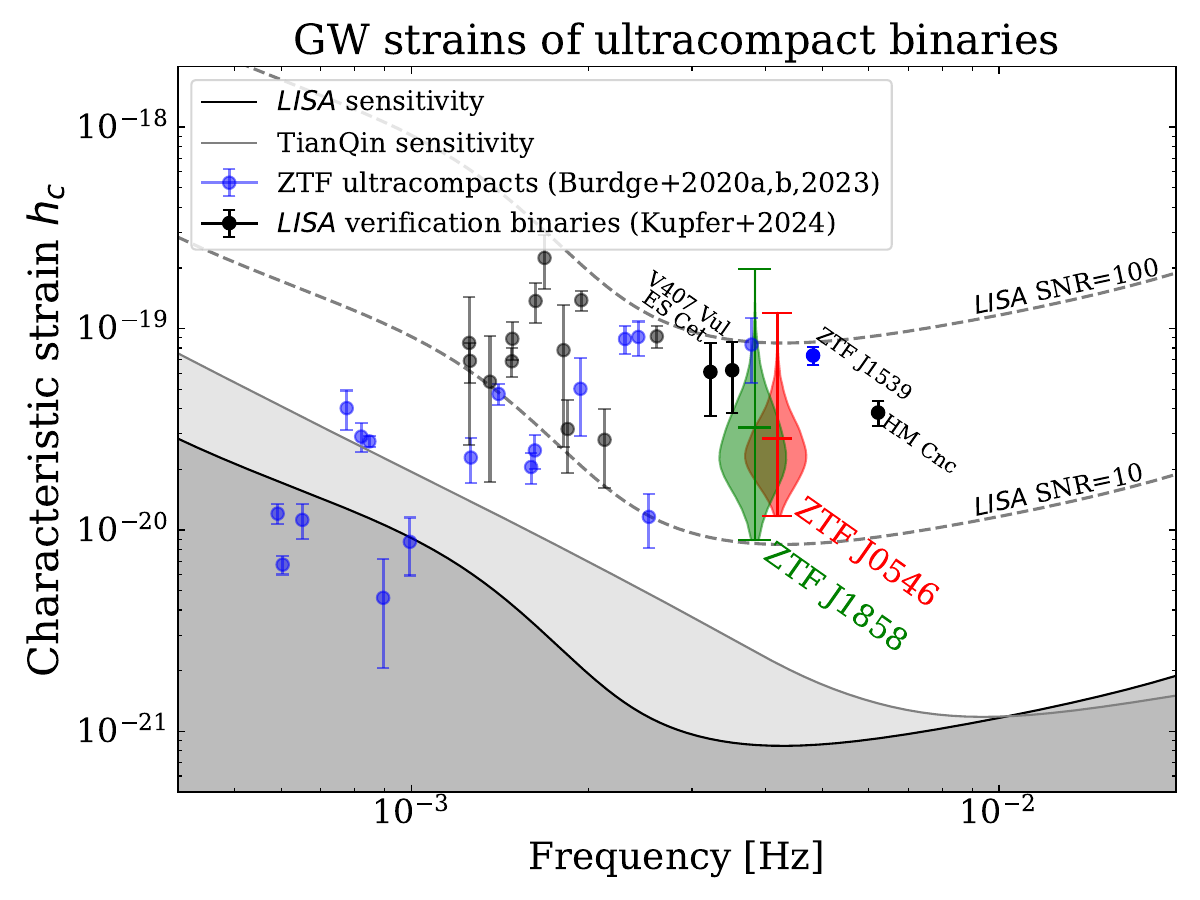}
    \caption{Violin plots of characteristic strains (over 4 yr) of ZTF J0546+3843 and ZTF J1858--2024 using the constraints from Section~\ref{subsec:pdot_constraints}. We also plot the strains of several ZTF-discovered ultracompact binaries \citep{Burdge2020a,Burdge2020b,Burdge2023} and \textit{LISA} verification binaries \citep{Kupfer2024} for comparison.}
    \label{fig:LISA_SNR}
\end{figure}

With estimates in hand for $M_1$, $M_2$ (Sec.~\ref{subsec:pdot_constraints}), and distance (Table~\ref{tab:objects}), we can readily assess the strength of the GW emission from these systems and ask: will these sources be detectable by space-based gravitational wave interformeters such as \textit{LISA}, TianQin, and Taiji, opening up the possibility of multi-messenger study? We show in Fig.~\ref{fig:LISA_SNR} that the answer to this question is a resounding yes. Using our $M_1$, $M_2$ distributions from Fig.~\ref{fig:pdot_constraints}, and drawing from a uniform distance distribution enclosing the 1-sigma posteriors (Table~\ref{tab:objects}), we computed characteristic GW strains assuming four-year missions. ZTF J0546+3843 and ZTF J1858--2024 are comfortably detected, with signal-to-noise ratios $>10$ even in pessimistic scenarios. 

This prompts a discussion of what additional insights are to be gained by supplementing electromagnetic (EM) observations of ultracompact binaries, such as those presented in this paper, with a GW signal. Significant previous work has been done examining the population-level inferences made possible from a multi-messenger signal (e.g. \citealt{Nelemans2004,Shah2012,Kupfer2018}); here, we instead focus on how joint EM/GW observations can aid in the detailed characterization of individual systems. We first review the basic observables of a GW signal, then outline how they can allow deeper constraints on these systems via the three key observables of (i) orbital phase, (ii) inclination, and (iii) chirp mass.

For our simplified case of a non-eccentric, slowly-evolving binary, the polarization amplitude of a GW signal is characterized by six parameters: distance ($d$), chirp mass ($\mathcal{M}_c$), two angles related to the sky position (RA $\alpha$, Dec $\delta$), and two angles related to the orbital plane orientation (inclination $\iota$, polarization angle $\psi$). The instantaneous GW strain in the source frame is a combination of three of the above parameters, and can be decomposed into two different polarizations:
\begin{equation}
    h_+(t) = \mathcal{A}\frac{1 + \cos^2 \iota}{2} \cos\bigg(\frac{4\pi t}{P} - \frac{\pi \dot{P}t^2}{P^2} + \phi_0\bigg)
    \label{eq:strain_plus}
\end{equation}
\begin{equation}
    h_\times(t) = \mathcal{A}\cos \iota \sin\bigg(\frac{4\pi t}{P} - \frac{\pi \dot{P}t^2}{P^2} + \phi_0\bigg)
    \label{eq:strain_cross}
\end{equation}
where $d$ and $\mathcal{M}_c$ are combined into the amplitude $\mathcal{A}$:
\begin{equation}
    \mathcal{A} = \frac{4(G\mathcal{M}_c)^{5/3}}{c^4 d}\bigg(\frac{2\pi}{P}\bigg)^{2/3}
    \label{eq:gw_amplitude}
\end{equation}
\citep{Shah2012}. The amplitude of the measured GW signal thus encodes a combination of the chirp mass, distance, and inclination. The remaining parameters enter when converting from the source frame to the radiation frame ($\psi$) and calculating the response of the detector for a given sky position ($\alpha$, $\delta$), but they are not intrinsically related to the system.

\begin{figure}
    \centering
    \includegraphics[width=\linewidth]{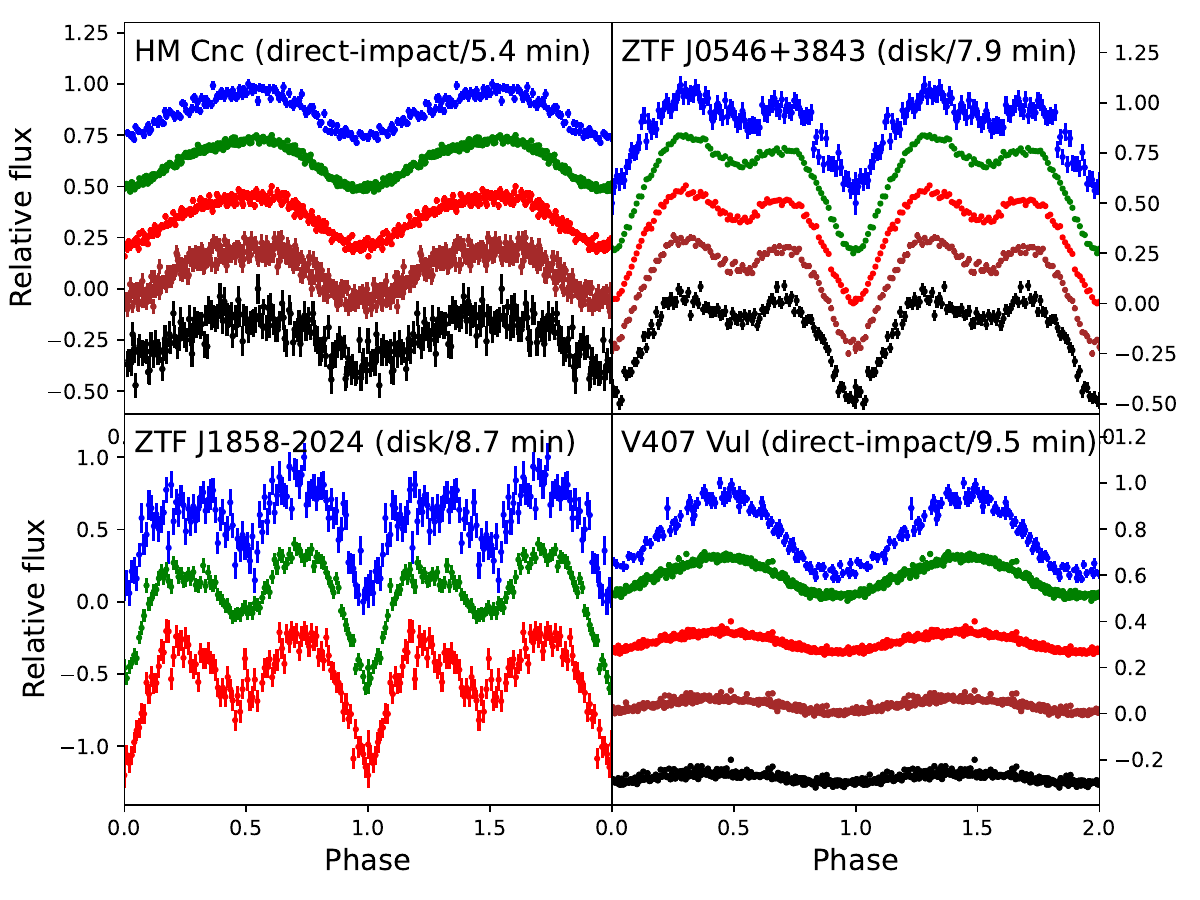}
    \caption{Multi-band phase-folded light curves of four illustrative ultracompact binaries ($u_s$/$g_s$/$r_s$/$i_s$/$z_s$ filters from top to bottom). Light curves of HM Cnc/V407 Vul are drawn from \cite{Munday2023}/\cite{MundayInPrep} respectively.}
    \label{fig:phase_fold_comparisons}
\end{figure}

\textbf{Orbital phase:} As the typical $\dot{P}$ in these systems is of $\mathcal{O}(10^{-11})$, their periods will change by only milliseconds over the course of the \textit{LISA} mission. For this reason they are often referred to as monochromatic GW sources, characterized by sinusoids of near-constant frequency. The relative simplicity of this signal is actually a blessing when taken alongside the far more complicated and uncertain EM signal. To illustrate this, in Fig.~\ref{fig:phase_fold_comparisons} we showcase the remarkable diversity of UV/optical/infrared light curves of four ultracompact binaries. Some systems show multiple eclipses interspersed with flicker noise from the accretion disk, while others show a quasi-sinusoidal waveform. Some show clear color-dependence in their variation indicating components of differing temperature, while others show almost an almost unchanged modulation across different wavelengths. Some are dominated by accretion luminosity, while others show a complicated blend of the donor/accretor stars alongside the signal from the accreting material. It is often impossible to assign a unique physical interpretation to each of these features with only EM data due to the many degrees of freedom, and the emergent huge variety of light curves, associated with accreting systems. As discussed in Sec.~\ref{subsec:optical}, this limits our ability to directly infer donor/accretor masses from radial velocity measurements, which can otherwise be done readily for detached systems.

However, a GW signal will uniquely determine the binary orbital phase via the right-most terms in Eqs. (\ref{eq:strain_plus}), (\ref{eq:strain_cross}). This is the single most constraining tool in interpreting the EM data: with complete phase information, one can uniquely determine which eclipses are of the donor, accretor, or accretion stream. This unlocks the potential of RV measurements to constrain masses, and in-eclipse color variations to constrain temperatures. With additional knowledge of the inclination one could also use the eclipses to measure donor/accretor radii; however, inclinations are difficult to constrain from EM data alone. Conveniently, a GW signal can also constrain this parameter.

\textbf{Inclination:} a GW detector is does not independently measure the polarizations $h_+$ and $h_\times$ directly. Rather, the detector response is set by a weighted sum of the two, with coefficients of the antenna beam patterns $F^+$, $F^\times$ \citep{Cornish2003}:
\begin{equation}
    h^2(t) \propto [h_+F^+(t,\alpha,\delta,\psi)]^2 + [h_\times F^\times(t,\alpha,\delta,\psi)]^2
\end{equation}
$F^+$ and $F^\times$ are time-dependent functions with a period of one year (for detectors co-orbiting with the Earth). The variation in $h(t)$ over a year of continuous observations can thus distinguish between systems of differing inclination.

For edge-on systems, a GW signal can constrain inclinations to within $\lesssim$10$^\circ$, which can then be combined with the photometric variation to measure donor/accretor radii via the eclipse depth. For all systems, inclinations together with complete phase information will unlock the power of RV measurements to constrain the component masses. The caveat is that not all systems will have visible donors/accretors, such as the ones in this work. But even for these systems, measurement of a chirp mass combined with our arguments in Section~\ref{subsec:pdot_constraints} can yield finer $M_1$/$M_2$ constraints, as we now discuss.

\textbf{Chirp mass:} a GW signal does not yield separate component masses, but their combination via the chirp mass, then too degenerate with the distance (Eq.~\ref{eq:gw_amplitude}). For sources with well-constrained EM distances, the latter degeneracy is lifted. Then, combining with our mass transfer arguments outlined in Section \ref{subsec:pdot_constraints} significantly constrains $M_1$ and $M_2$: only a small range of solutions lying within the density contours of Fig.~\ref{fig:pdot_constraints} will agree with a measured $\mathcal{M}_c$. Coupled with radius constrains as discussed above, joint EM/GW measurements will directly constrain the amount of entropy in the donor WDs via their mass-radius relations, which is a key uncertainty in current binary evolution models \citep{Wong2021}.

With the precise mass constraints unlocked by multi-messenger study of accreting binaries, one can compare an observed Type Ia SNe rate to the incidence of ultracompact binaries with sufficient mass to induce a double-degenerate supernova \citep{Shen2015}. Adding the radius constraints further probes the white dwarf/He star mass-radius relation (Sec.~\ref{subsec:pdot_constraints}) which is set by the amount of residual thermal energy remaining from the common envelope phase. Finally, population studies of accreting systems across a range of inclinations will be a direct probe into the vertical structure of accretion disks, whose scale heights as a function of accretion rate and radial extent are highly uncertain.

\subsection{Orbital Evolution of the Ultracompact Binary Population}

\begin{table*}
\centering
\begin{tabular}{c|c|c|c|c|c}
Name & $P$ (min) & $\dot{P}$ (s s$^{-1}$)& $\mathcal{M}_c$ ($M_\odot$) & System type & References \\
\hline \hline
HM Cnc & 5.35 & $(-3.66\pm 0.001)\times 10^{-11}$ & --- & Direct-impact accretor & 1,2,3 \\
eRASSU J060839.5-704014 & 6.23 & --- & --- & Likely direct-impact accretor & 4  \\
ZTF J1539+5027 & 6.91 & $(-2.37\pm 0.005) \times 10^{-11}$ & $0.30\pm 0.01$ & Detached binary & 5 \\
\textbf{ZTF J0546+3843} & 7.95 & $(-4.30^{+1.1}_{-1.0}) \times 10^{-12}$ & --- & Disk accretor & \textbf{This work} \\
\textbf{ZTF J1858--2024} & 8.68 & $(+7.80^{+1.70}_{-1.12}) \times 10^{-12}$ & --- & Disk accretor & \textbf{This work} \\
ZTF J2243+5242 &  8.80 & $(-1.6 \pm 0.18) \times 10^{-11}$ & $0.31 \pm 0.06$ & Detached binary & 6 \\
V407 Vul & 9.48 & $(-2.27 \pm 0.25) \times 10^{-12}$ & --- & Direct-impact accretor & 7,8,9 \\
ES Cet & 10.33 & $(+3.2 \pm 0.1)\times 10^{-12}$ & --- & Disk accretor & 10,11 \\
SDSS J0651+2844 & 12.75 & $(-9.8 \pm 2.8) \times 10^{-12}$ & $0.30 \pm 0.01$ & Detached binary & 12 \\
\textbf{ZTF J0425+3858} & 13.15 & --- & --- & Disk accretor & \textbf{This work} \\
ZTF J0127+5258 & 13.71 & $(-6.53 \pm 0.19) \times 10^{-12}$ & $0.31 \pm 0.03$ & Disk accretor & 13 \\
\end{tabular}
\\
\caption{Orbital period ($P$), period derivative ($\dot{P}$), and chirp mass ($\mathcal{M}_c$) for the shortest-period binaries systems known. Chirp masses are only shown for systems where $M_1$ and $M_2$ are directly measured. \textit{References:} (1) \citealt{Roelofs2010} (2) \citealt{Strohmayer2021} (3) \citealt{Munday2023} (4) \citealt{Maitra2024} (5) \citealt{Burdge2019} (6) \citealt{Burdge2020b} (7) \citealt{Motch1996} (8) \citealt{Marsh2002} (9) \citealt{Strohmayer2004a} (10) \citealt{Konda1984} (11) \citealt{deMiguel2018} (12) \citealt{Hermes2012} (13) \citealt{Burdge2023}} \label{tab:p_pdots}
\end{table*}

Having significantly expanded the sample of accreting ultracompacts with measured $\dot{P}$ (Table~\ref{tab:p_pdots}), we now discuss population-level implications for the trends observed so far in the orbital evolution of the shortest-period accreting binaries.

\begin{figure}
    \centering
    \includegraphics[width=\linewidth]{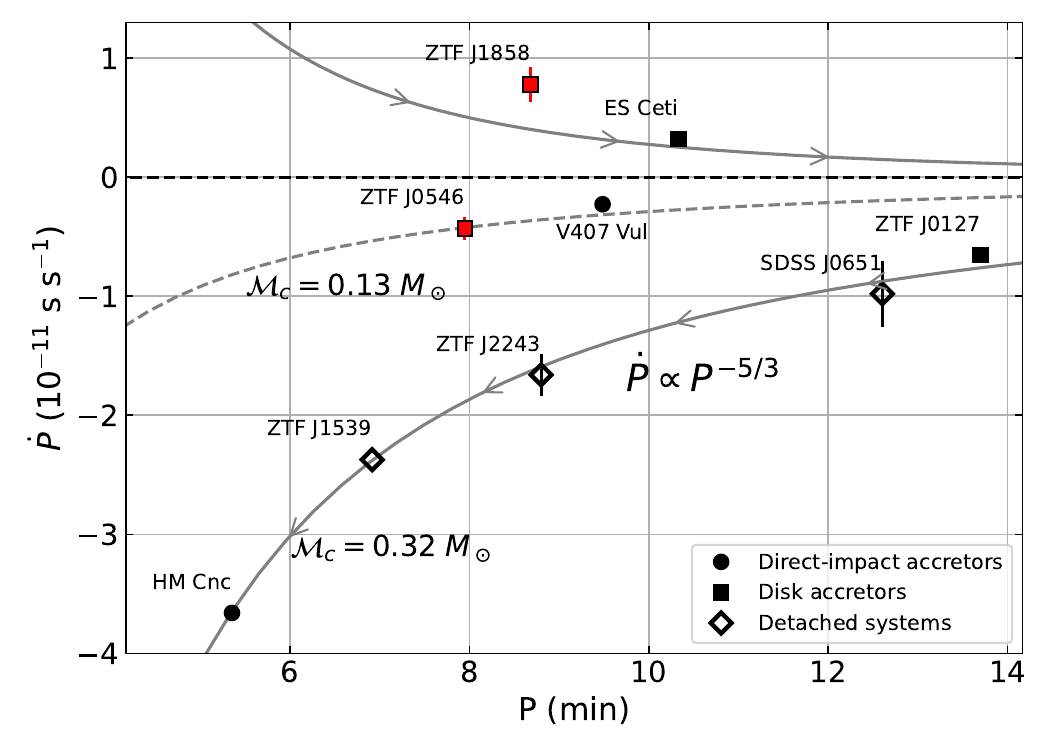}
    \caption{The orbital period and its derivative ($\dot{P}$) for all ultracompact binaries in which both are measured (Table~\ref{tab:p_pdots}). Our new systems are highlighted in red. Pure gravitational-wave decay, in the absence of mass transfer and tides, corresponds to $\dot{P} \propto P^{-5/3}$, which we illustrate for two different chirp masses. All detached systems lie nearly on the same curve of constant chirp mass ($\mathcal{M}_c=0.32\;M_\odot$), while some accreting systems diverge significantly due to the effects of mass transfer.}
    \label{fig:p_pdot_tracks}
\end{figure}

It is of great relevance whether initially detached systems, shortly after they come into mass-transfer contact, can stabilize via efficient angular momentum transfer back to the orbit. The generally assumed mechanism for this angular momentum transfer is synchronization of the accretor spin with the binary orbit via tidal torques \citep{Marsh2004,Gokhale2007}. The fraction of systems which stabilize vs. merge is an extremely sensitive function of the timescale over which this angular momentum redistribution occurs (the tidal synchronization timescale; \citealt{Kremer2015,Biscoveanu2023,Toubiana2024}). Numerical simulations in \cite{Marsh2004} note that this timescale needs to be extremely rapid, $\mathcal{O}(1000)$ yr, for a significant fraction of systems to stabilize. 

\cite{Strohmayer2021} and \cite{Munday2023} reported that HM Cnc, despite having a negative $\dot{P}$ as expected from GR-dominated decay, has a positive second derivative ($\ddot{P}$); the inferred period minimum will occur within $\sim 1200-2000$ yr, after which HM Cnc will begin to outspiral. The fact that HM Cnc was caught a mere $\lesssim 2000$ years prior to period minimum suggests that there is a selection bias at play, namely that $\dot{M}$ and $L$ peaks enormously during the short-lived passage through period minimum, making the systems much easier to discover. This is broadly in agreement with evolutionary models and simulations, which find $\dot{M}$ spikes of 2-3 orders of magnitude near turnaround \citep{Marsh2004,Motl2007}. The lower observed rate of direct-impact accretors compared to detached and disk-accreting systems (Table~\ref{tab:p_pdots}) is further evidence towards this speculation, which will be tested by discovery of further systems at the shortest periods. Sufficiently long-baseline timing observations of ZTF J0546+3843 will also allow direct detection of a $\ddot{P}$ (which for the derived mass constraints in Section~\ref{subsec:pdot_constraints} is likely negative). Comparison of a measured $\ddot{P}$ with that of HM Cnc would constitute an observational test of the efficiency of tidal torques with and without an accretion disk, with implications for the total number of direct- vs. disk-accreting systems in the Galaxy at a given time.

The remarkable accuracy with which several inspiraling systems follow a single $P^{-5/3}$ curve (Fig.~\ref{fig:p_pdot_tracks}), across both accreting and detached systems, suggests comparable chirp masses ($\mathcal{M}_c$) in all systems, suggesting a common binary evolution process is responsible for the majority of stable systems at these orbital separations. Indeed, in the systems with direct mass measurements (mostly detached binaries), all have $\mathcal{M}_c\approx 0.3\;M_\odot$ (Table~\ref{tab:p_pdots}). The notable exceptions from this curve are V407 Vul and ZTF J0546+3843, with $|\dot{P}|$ several times smaller than naively anticipated from the empirical trend in Fig.~\ref{fig:p_pdot_tracks}. The two possibilities are that (1) the chirp mass in these systems is smaller, either via the effect of sustained mass transfer or an alternate formation channel; or (2) these systems are on course for a period minimum ($\dot{P}=0$) within the next few thousand of years. Case (2) would be additional evidence for efficient tidal synchronization in ZTF J0546+3843, as this implies a large $\ddot{P}$ (hence a large spread in $\dot{P}$). Measurements of $\dot{P}$ in more accreting systems, or directly constraining $\ddot{P}$ in V407 Vul/ZTF J0546+3843, would test this hypothesis.

We also consider the relative likelihood of detecting systems near $P_{\mathrm{min}}$ in their inspiral vs. outspiral phase. The CO + ELM WD evolutionary models of \cite{Kaplan2012} found that, as the time spent at a particular period scales as $\propto |1/\dot{P}|$, the model population of ingoing vs. outgoing systems should be:
\begin{linenomath*}
\[
\mathcal{N}_{\mathrm{out/in}} \equiv \bigg|\frac{\dot{P}_{\mathrm{in}}}{\dot{P}_{\mathrm{out}}}\bigg| \approx 4\bigg(\frac{P}{10\;\mathrm{min}}\bigg)\bigg(\frac{M_{2,i}}{0.15 M_\odot}\bigg)
\]
\end{linenomath*}
Considering only the systems with donor/accretor masses near 0.8/0.15 $M_\odot$ (so that the evolutionary models of \citealt{Kaplan2012} are relevant), the observed ratio ingoing vs.~outgoing systems $\mathcal{N}_{\mathrm{out/in}} = 2/4$ (Table~\ref{tab:p_pdots}) appears in tension with the results of these evolutionary models, albeit still with small-number statistics. There are a handful of possible unmodelled effects contributing to this disagreement. For one, most systems may never stabilize to produce a long-lived outspiraling AM CVn due to inefficient tidal synchronization or nova eruptions resulting in rapid orbital destablization and merger \citep{Shen2015}. Another alternative is that our understanding of the outspiraling phase is incomplete, and $\dot{P}_{\mathrm{out}}$ is significantly larger than expected for unknown reasons. The larger-than-expected $\dot{P}$ of ZTF J1858--2024 provides compelling evidence towards this possibility.

As a large enough sample of $\dot{P}$ measurements for population studies is just beginning to emerge, we cannot favor either scenario with confidence, but time will tell whether our theoretical picture of double WDs near period minimum requires significant rethinking. This work represents an important step towards building up the statistics to rigorously test these evolution models.

\section{Conclusions} \label{sec:conclusion}

In this work, we present the discovery of three new ultracompact accreting binaries, with orbital periods of 7.95, 8.68, and 13.15 minutes. We draw the following conclusions:
\begin{itemize}
    \item We infer that all three systems accrete via a disk from the presence of double-peaked emission lines in their spectra (Figs.~\ref{fig:spec} \& \ref{fig:trail}). Previously only direct-impact accreting systems were known below 10 minutes (Table~\ref{tab:p_pdots}), but our new systems show that the accretors in ultracompact binaries can get compact enough to host accretion disks even at these periods.
    \item All three systems show remarkably similar optical spectra dominated by ionized helium and nitrogen (Fig.~\ref{fig:spec}). The similarity of the outspiraling systems ZTF J1858--2024 and ES Cet with the inspiraling binary ZTF J0546+3843---especially the lack of hydrogen in the latter---is particularly surprising from an evolutionary standpoint. We obtained an FUV spectrum indicating lower N/C in ZTF J0546+3843 compared to ES Cet (Fig.~\ref{fig:UV}), which is qualitatively consistent with a less-stripped donor and/or a less-massive main sequence progenitor. 
    \item We measured ZTF J0546+3843 to have $\dot{P}=-4.30^{+1.1}_{-1.0} \times 10^{-12}$ s s$^{-1}$, and ZTF J1858--2024 to have $\dot{P}=+7.80^{+1.70}_{-1.12} \times 10^{-12}$ s s$^{-1}$ (Fig.~\ref{fig:timing}). We have shown how the measured $\dot{P}/P$ can be combined with mass-transfer arguments, as well as some loose assumptions about the adiabatic and thermal/nuclear evolution properties of the donor, to place tighter bounds on the system parameters (Fig.~\ref{fig:pdot_constraints}). With these constraints, we infer the donor in ZTF J1858--2024 to have an adiabatic index comparable to that expected for a zero-temperature degenerate star, while the donor in ZTF J0546+3843 must have $\xi_{\mathrm{ad}}$ significantly larger. The small magnitude of $\dot{P}$ in ZTF J0546+3843 heavily implies a system close to period minimum, whereas the large $\dot{P}$ in ZTF J1858--2024 is difficult to explain with current binary evolution models (Fig.~\ref{fig:mesa_tracks}).
    \item Based on the range of allowed donor/accretor masses and distance posteriors, we infer that ZTF J0546+3843 and ZTF J1858--2024 are loud GW sources detectable by future space-based millihertz GW detectors, e.g. TianQin, Taiji, and \textit{LISA} (Fig.~\ref{fig:LISA_SNR}). A GW signal, when combined with our EM analysis and mass-transfer arguments, can significantly constrain the component masses/radii, system inclinations, and donor structure (Sec.~\ref{subsec:multimessenger}), demonstrating a compelling multi-messenger science case.
    \item At a population level, the observed scarcity of direct-impact accretors, compared to disk-accretors and detached systems at short orbital periods (Table~\ref{tab:p_pdots}) seems to suggest there is an observational bias against detecting them, e.g. if their high-$\dot{M}$ phase is very short-lived. Such an effect could be achieved if the tidal synchronization efficiency scales sharply with $\dot{M}$, so that period turnaround occurs rapidly after $\dot{M}$ spikes. A future measurement of $\ddot{P}$ in the inspiraling system ZTF J0546+3843, compared to the recent positive $\ddot{P}$ measured for HM Cnc \citep{Strohmayer2021,Munday2023}, will allow directly comparing the efficiency of tidal synchronization with and without an accretion disk. It is also curious that only two outspiraling systems are known, while the population of inspiraling systems seemingly continues to grow. It is possible there is an observational selection effect at play, or that most interacting double WDs are unstable to mergers, lowering the rate of long-lived AM CVn \citep{Shen2015}. The implications of these effects will extend to Type Ia SNa rates \citep{Ruiter2009}, galaxy-scale binary population synthesis models \citep{Breivik2020}, and the all-sky mHz GW signal \citep{Littenberg2020}. 
\end{itemize}

\section*{Acknowledgements}
We thank the anonymous referee for useful comments which improved the quality of the paper. We thank Karolina Bąkowska for providing us the optical spectrum of ES Cet \citep{Bakowska2021}. We also thank Jim Fuller and Antonio Rodriguez for interesting discussions regarding this work.

PR--G acknowledges support by the Spanish Agencia Estatal de Investigación del Ministerio de Ciencia e Innovación (MCIN/AEI) and the European Regional Development Fund (ERDF) under grant PID2021--124879NB--I00. IP acknowledges support from a Royal Society University Research Fellowship (URF\textbackslash R1\textbackslash231496). H-LC is supported by the National Key R\&D Program of China (grant Nos. 2021YFA1600403), the National Natural Science Foundation of China (grant Nos. 12288102, 12333008 and 12422305).

Based on observations obtained with the Samuel Oschin Telescope 48-inch and the 60-inch Telescope at the Palomar Observatory as part of the Zwicky Transient Facility project. ZTF is supported by the National Science Foundation under Grant No. AST-1440341 and a collaboration including Caltech, IPAC, the Weizmann Institute for Science, the Oskar Klein Center at Stockholm University, the University of Maryland, the University of Washington, Deutsches ElektronenSynchrotron and Humboldt University, Los Alamos National Laboratories, the TANGO Consortium of Taiwan,
the University of Wisconsin at Milwaukee, and Lawrence Berkeley National Laboratories. Operations are conducted by COO, IPAC, and UW.

Based on observations made with the Gran Telescopio Canarias (GTC) installed in the Spanish Observatorio del Roque de los Muchachos of the Instituto de Astrofísica de Canarias, on the island of La Palma. Based on observations made at the European Southern Observatory New Technology Telescope (NTT), La Silla.

VSD, ULTRACAM and HiPERCAM are funded by the Science and Technology Facilities Council (grant ST/Z000033/1).

The KPED team thanks the National Science Foundation and the National Optical Astronomical Observatory for making the Kitt Peak 2.1-m telescope available. The
KPED team thanks the National Science Foundation, the National Optical Astronomical Observatory and the Murty family for support in the building and operation of KPED. In addition, they thank the CHIMERA project for use of the Electron Multiplying CCD (EMCCD). Some of the data presented herein were obtained at the W.M. Keck Observatory, which is operated as a scientific partnership among the California Institute of Technology, the University of California and the National Aeronautics and Space  Administration. The Observatory was made possible by the generous financial support of the W.~M. Keck Foundation. The authors wish to recognize and acknowledge the very significant cultural role and reverence that the summit of Mauna Kea has always had within the indigenous Hawaiian community. We are most fortunate to have the opportunity to conduct observations from this mountain.

\appendix

\setcounter{table}{0}
\renewcommand{\thetable}{A.\arabic{table}} 
\section{Supplementary Figures and Tables}
\begin{table}[h]
    \centering
    \caption{Observation log of all high-speed photometry epochs. These observations were fit with a quadratic timing model to determine the $\dot{P}$ as shown in Fig.~\ref{fig:timing}.}
    \label{tab:observing_log}
    \begin{tabular}{l|c|c|c|c|c|c}
        Object & Instrument & Date & Filters & Total Exposure & Exposure Time & $T_0$ \\
        & & & & [hr] & [s] & [BJD] \\
        \hline \hline
        \multirow{8}{*}{ZTF J0546+3843} 
        & CHIMERA & 2020-12-15 & g/r & 8.25 & 6 & 59198.23459(7) \\
        & CHIMERA & 2021-04-13 & g/r & 1.84 & 11 & 59317.16810(6)  \\
        & CHIMERA & 2021-10-10 & g/r & 2.01 & 5 & 59497.38655(8) \\
        & CHIMERA & 2022-08-25 & g/r & 1.15 & 5 & 59816.47734(4) \\
        & HiPERCAM & 2024-03-10 & u$_s$/g$_s$/r$_s$/i$_s$/z$_s$ & 2.85 & 1.7 & 60379.85838(2) \\
        & HiPERCAM & 2024-08-28 & u$_s$/g$_s$/r$_s$/i$_s$/z$_s$ & 0.47 & 1.7 & 60550.20934(3) \\
        & HiPERCAM & 2024-08-29 & u$_s$/g$_s$/r$_s$/i$_s$/z$_s$ & 0.95 & 1.7 & 60551.18058(3) \\
        & HiPERCAM & 2024-08-31 & u$_s$/g$_s$/r$_s$/i$_s$/z$_s$ & 0.47 & 1.7 & 60553.18938(1) \\
        & Lightspeed & 2024-10-26 & L & 0.60 & 1.0 & 60609.39173(5) \\
        \hline
        \multirow{7}{*}{ZTF J1858--2024} 
        & CHIMERA & 2019-04-28 & g/r & 1.67 & 3.0 & 58601.4689(8) \\
        & ULTRACAM & 2019-07-09 & g$_s$/r$_s$/i$_s$ & 1.02 & 6.5 & 58673.3068(2) \\
        & ULTRACAM & 2019-07-10 & g$_s$/r$_s$/i$_s$ & 2.01 & 6.5 & 58674.1387(4)  \\
        & CHIMERA & 2019-10-01 & g/r & 1.22 & 5.5 & 58757.1164(8) \\
        & ULTRACAM & 2023-04-01 & g$_s$/r$_s$/i$_s$ & 1.22 & 5.7 & 60035.3702(3) \\
        & HiPERCAM & 2024-05-22 & u$_s$/g$_s$/r$_s$/i$_s$/z$_s$ & 2.68 & 3.8 & 60452.1204(1) \\
        & ULTRACAM & 2024-07-30 & g$_s$/r$_s$/i$_s$ & 1.18 & 10 & 60521.0415(2) \\
        \hline
        \multirow{1}{*}{ZTF J0425+3858} 
        & CHIMERA & 2022-08-24 & g/r & 0.84 & 5.63 & 59815.4026(4)
    \end{tabular}
\end{table}

\bibliography{refs}{}

\begin{thebibliography}{}
\expandafter\ifx\csname natexlab\endcsname\relax\def\natexlab#1{#1}\fi
\providecommand{\url}[1]{\href{#1}{#1}}
\providecommand{\dodoi}[1]{doi:~\href{http://doi.org/#1}{\nolinkurl{#1}}}
\providecommand{\doeprint}[1]{\href{http://ascl.net/#1}{\nolinkurl{http://ascl.net/#1}}}
\providecommand{\doarXiv}[1]{\href{https://arxiv.org/abs/#1}{\nolinkurl{https://arxiv.org/abs/#1}}}

\bibitem[{{Amaro-Seoane} {et~al.}(2017){Amaro-Seoane}, {Audley}, {Babak}, {Baker}, {Barausse}, {Bender}, {Berti}, {Binetruy}, {Born}, {Bortoluzzi}, {Camp}, {Caprini}, {Cardoso}, {Colpi}, {Conklin}, {Cornish}, {Cutler}, {Danzmann}, {Dolesi}, {Ferraioli}, {Ferroni}, {Fitzsimons}, {Gair}, {Gesa Bote}, {Giardini}, {Gibert}, {Grimani}, {Halloin}, {Heinzel}, {Hertog}, {Hewitson}, {Holley-Bockelmann}, {Hollington}, {Hueller}, {Inchauspe}, {Jetzer}, {Karnesis}, {Killow}, {Klein}, {Klipstein}, {Korsakova}, {Larson}, {Livas}, {Lloro}, {Man}, {Mance}, {Martino}, {Mateos}, {McKenzie}, {McWilliams}, {Miller}, {Mueller}, {Nardini}, {Nelemans}, {Nofrarias}, {Petiteau}, {Pivato}, {Plagnol}, {Porter}, {Reiche}, {Robertson}, {Robertson}, {Rossi}, {Russano}, {Schutz}, {Sesana}, {Shoemaker}, {Slutsky}, {Sopuerta}, {Sumner}, {Tamanini}, {Thorpe}, {Troebs}, {Vallisneri}, {Vecchio}, {Vetrugno}, {Vitale}, {Volonteri}, {Wanner}, {Ward}, {Wass}, {Weber}, {Ziemer}, \& {Zweifel}}]{Amaro-Seoane2017}
{Amaro-Seoane}, P., {Audley}, H., {Babak}, S., {et~al.} 2017, arXiv e-prints, arXiv:1702.00786.
\newblock \doarXiv{1702.00786}

\bibitem[{{Amaro-Seoane} {et~al.}(2023){Amaro-Seoane}, {Andrews}, {Arca Sedda}, {Askar}, {Baghi}, {Balasov}, {Bartos}, {Bavera}, {Bellovary}, {Berry}, {Berti}, {Bianchi}, {Blecha}, {Blondin}, {Bogdanovi{\'c}}, {Boissier}, {Bonetti}, {Bonoli}, {Bortolas}, {Breivik}, {Capelo}, {Caramete}, {Cattorini}, {Charisi}, {Chaty}, {Chen}, {Chru{\'s}li{\'n}ska}, {Chua}, {Church}, {Colpi}, {D'Orazio}, {Danielski}, {Davies}, {Dayal}, {De Rosa}, {Derdzinski}, {Destounis}, {Dotti}, {Du{\c{t}}an}, {Dvorkin}, {Fabj}, {Foglizzo}, {Ford}, {Fouvry}, {Franchini}, {Fragos}, {Fryer}, {Gaspari}, {Gerosa}, {Graziani}, {Groot}, {Habouzit}, {Haggard}, {Haiman}, {Han}, {Istrate}, {Johansson}, {Khan}, {Kimpson}, {Kokkotas}, {Kong}, {Korol}, {Kremer}, {Kupfer}, {Lamberts}, {Larson}, {Lau}, {Liu}, {Lloyd-Ronning}, {Lodato}, {Lupi}, {Ma}, {Maccarone}, {Mandel}, {Mangiagli}, {Mapelli}, {Mathis}, {Mayer}, {McGee}, {McKernan}, {Miller}, {Mota}, {Mumpower}, {Nasim}, {Nelemans}, {Noble}, {Pacucci}, {Panessa}, {Paschalidis}, {Pfister}, {Porquet},
  {Quenby}, {Ricarte}, {R{\"o}pke}, {Regan}, {Rosswog}, {Ruiter}, {Ruiz}, {Runnoe}, {Schneider}, {Schnittman}, {Secunda}, {Sesana}, {Seto}, {Shao}, {Shapiro}, {Sopuerta}, {Stone}, {Suvorov}, {Tamanini}, {Tamfal}, {Tauris}, {Temmink}, {Tomsick}, {Toonen}, {Torres-Orjuela}, {Toscani}, {Tsokaros}, {Unal}, {V{\'a}zquez-Aceves}, {Valiante}, {van Putten}, {van Roestel}, {Vignali}, {Volonteri}, {Wu}, {Younsi}, {Yu}, {Zane}, {Zwick}, {Antonini}, {Baibhav}, {Barausse}, {Bonilla Rivera}, {Branchesi}, {Branduardi-Raymont}, {Burdge}, {Chakraborty}, {Cuadra}, {Dage}, {Davis}, {de Mink}, {Decarli}, {Doneva}, {Escoffier}, {Gandhi}, {Haardt}, {Lousto}, {Nissanke}, {Nordhaus}, {O'Shaughnessy}, {Portegies Zwart}, {Pound}, {Schussler}, {Sergijenko}, {Spallicci}, {Vernieri}, \& {Vigna-G{\'o}mez}}]{Amaro-Seoane2023}
{Amaro-Seoane}, P., {Andrews}, J., {Arca Sedda}, M., {et~al.} 2023, Living Reviews in Relativity, 26, 2, \dodoi{10.1007/s41114-022-00041-y}

\bibitem[{{Bailer-Jones} {et~al.}(2021){Bailer-Jones}, {Rybizki}, {Fouesneau}, {Demleitner}, \& {Andrae}}]{BailerJones2021}
{Bailer-Jones}, C.~A.~L., {Rybizki}, J., {Fouesneau}, M., {Demleitner}, M., \& {Andrae}, R. 2021, \aj, 161, 147, \dodoi{10.3847/1538-3881/abd806}

\bibitem[{{Begari} \& {Maccarone}(2023)}]{Begari2023}
{Begari}, T., \& {Maccarone}, T.~J. 2023, \jaavso, 51, 227, \dodoi{10.48550/arXiv.2312.06007}

\bibitem[{{Bellm} {et~al.}(2019){Bellm}, {Kulkarni}, {Graham}, {Dekany}, {Smith}, {Riddle}, {Masci}, {Helou}, {Prince}, {Adams}, {Barbarino}, {Barlow}, {Bauer}, {Beck}, {Belicki}, {Biswas}, {Blagorodnova}, {Bodewits}, {Bolin}, {Brinnel}, {Brooke}, {Bue}, {Bulla}, {Burruss}, {Cenko}, {Chang}, {Connolly}, {Coughlin}, {Cromer}, {Cunningham}, {De}, {Delacroix}, {Desai}, {Duev}, {Eadie}, {Farnham}, {Feeney}, {Feindt}, {Flynn}, {Franckowiak}, {Frederick}, {Fremling}, {Gal-Yam}, {Gezari}, {Giomi}, {Goldstein}, {Golkhou}, {Goobar}, {Groom}, {Hacopians}, {Hale}, {Henning}, {Ho}, {Hover}, {Howell}, {Hung}, {Huppenkothen}, {Imel}, {Ip}, {Ivezi{\'c}}, {Jackson}, {Jones}, {Juric}, {Kasliwal}, {Kaspi}, {Kaye}, {Kelley}, {Kowalski}, {Kramer}, {Kupfer}, {Landry}, {Laher}, {Lee}, {Lin}, {Lin}, {Lunnan}, {Giomi}, {Mahabal}, {Mao}, {Miller}, {Monkewitz}, {Murphy}, {Ngeow}, {Nordin}, {Nugent}, {Ofek}, {Patterson}, {Penprase}, {Porter}, {Rauch}, {Rebbapragada}, {Reiley}, {Rigault}, {Rodriguez}, {van Roestel}, {Rusholme}, {van
  Santen}, {Schulze}, {Shupe}, {Singer}, {Soumagnac}, {Stein}, {Surace}, {Sollerman}, {Szkody}, {Taddia}, {Terek}, {Van Sistine}, {van Velzen}, {Vestrand}, {Walters}, {Ward}, {Ye}, {Yu}, {Yan}, \& {Zolkower}}]{Bellm2019}
{Bellm}, E.~C., {Kulkarni}, S.~R., {Graham}, M.~J., {et~al.} 2019, \pasp, 131, 018002, \dodoi{10.1088/1538-3873/aaecbe}

\bibitem[{{Biscoveanu} {et~al.}(2023){Biscoveanu}, {Kremer}, \& {Thrane}}]{Biscoveanu2023}
{Biscoveanu}, S., {Kremer}, K., \& {Thrane}, E. 2023, \apj, 949, 95, \dodoi{10.3847/1538-4357/acc585}

\bibitem[{{B{\k{a}}kowska} {et~al.}(2021){B{\k{a}}kowska}, {Marsh}, {Steeghs}, {Nelemans}, \& {Groot}}]{Bakowska2021}
{B{\k{a}}kowska}, K., {Marsh}, T.~R., {Steeghs}, D., {Nelemans}, G., \& {Groot}, P.~J. 2021, \aap, 645, A114, \dodoi{10.1051/0004-6361/202039266}

\bibitem[{{Breivik} {et~al.}(2020){Breivik}, {Coughlin}, {Zevin}, {Rodriguez}, {Kremer}, {Ye}, {Andrews}, {Kurkowski}, {Digman}, {Larson}, \& {Rasio}}]{Breivik2020}
{Breivik}, K., {Coughlin}, S., {Zevin}, M., {et~al.} 2020, \apj, 898, 71, \dodoi{10.3847/1538-4357/ab9d85}

\bibitem[{{Buchner}(2021)}]{Buchner2021}
{Buchner}, J. 2021, The Journal of Open Source Software, 6, 3001, \dodoi{10.21105/joss.03001}

\bibitem[{{Burdge}(in prep., 2025)}]{BurdgeInPrep}
{Burdge}, K. in prep., 2025

\bibitem[{{Burdge} {et~al.}(2019){Burdge}, {Coughlin}, {Fuller}, {Kupfer}, {Bellm}, {Bildsten}, {Graham}, {Kaplan}, {Roestel}, {Dekany}, {Duev}, {Feeney}, {Giomi}, {Helou}, {Kaye}, {Laher}, {Mahabal}, {Masci}, {Riddle}, {Shupe}, {Soumagnac}, {Smith}, {Szkody}, {Walters}, {Kulkarni}, \& {Prince}}]{Burdge2019}
{Burdge}, K.~B., {Coughlin}, M.~W., {Fuller}, J., {et~al.} 2019, \nat, 571, 528, \dodoi{10.1038/s41586-019-1403-0}

\bibitem[{{Burdge} {et~al.}(2020{\natexlab{a}}){Burdge}, {Prince}, {Fuller}, {Kaplan}, {Marsh}, {Tremblay}, {Zhuang}, {Bellm}, {Caiazzo}, {Coughlin}, {Dhillon}, {Gaensicke}, {Rodr{\'\i}guez-Gil}, {Graham}, {Hermes}, {Kupfer}, {Littlefair}, {Mr{\'o}z}, {Phinney}, {van Roestel}, {Yao}, {Dekany}, {Drake}, {Duev}, {Hale}, {Feeney}, {Helou}, {Kaye}, {Mahabal}, {Masci}, {Riddle}, {Smith}, {Soumagnac}, \& {Kulkarni}}]{Burdge2020a}
{Burdge}, K.~B., {Prince}, T.~A., {Fuller}, J., {et~al.} 2020{\natexlab{a}}, \apj, 905, 32, \dodoi{10.3847/1538-4357/abc261}

\bibitem[{{Burdge} {et~al.}(2020{\natexlab{b}}){Burdge}, {Coughlin}, {Fuller}, {Kaplan}, {Kulkarni}, {Marsh}, {Bellm}, {Dekany}, {Duev}, {Graham}, {Mahabal}, {Masci}, {Laher}, {Riddle}, {Soumagnac}, \& {Prince}}]{Burdge2020b}
{Burdge}, K.~B., {Coughlin}, M.~W., {Fuller}, J., {et~al.} 2020{\natexlab{b}}, \apjl, 905, L7, \dodoi{10.3847/2041-8213/abca91}

\bibitem[{{Burdge} {et~al.}(2022){Burdge}, {El-Badry}, {Marsh}, {Rappaport}, {Brown}, {Caiazzo}, {Chakrabarty}, {Dhillon}, {Fuller}, {G{\"a}nsicke}, {Graham}, {Kara}, {Kulkarni}, {Littlefair}, {Mr{\'o}z}, {Rodr{\'\i}guez-Gil}, {Roestel}, {Simcoe}, {Bellm}, {Drake}, {Dekany}, {Groom}, {Laher}, {Masci}, {Riddle}, {Smith}, \& {Prince}}]{Burdge2022}
{Burdge}, K.~B., {El-Badry}, K., {Marsh}, T.~R., {et~al.} 2022, \nat, 610, 467, \dodoi{10.1038/s41586-022-05195-x}

\bibitem[{{Burdge} {et~al.}(2023){Burdge}, {El-Badry}, {Rappaport}, {Wong}, {Bauer}, {Bildsten}, {Caiazzo}, {Chakrabarty}, {Chickles}, {Graham}, {Kara}, {Kulkarni}, {Marsh}, {Nynka}, {Prince}, {Simcoe}, {van Roestel}, {Vanderbosch}, {Bellm}, {Dekany}, {Drake}, {Helou}, {Masci}, {Milburn}, {Riddle}, {Rusholme}, \& {Smith}}]{Burdge2023}
{Burdge}, K.~B., {El-Badry}, K., {Rappaport}, S., {et~al.} 2023, arXiv e-prints, arXiv:2303.13573, \dodoi{10.48550/arXiv.2303.13573}

\bibitem[{{Chen} {et~al.}(2022){Chen}, {Chen}, \& {Han}}]{Chen2022}
{Chen}, H.-L., {Chen}, X., \& {Han}, Z. 2022, \apj, 935, 9, \dodoi{10.3847/1538-4357/ac7fec}

\bibitem[{{Cornish} \& {Larson}(2003)}]{Cornish2003}
{Cornish}, N.~J., \& {Larson}, S.~L. 2003, Classical and Quantum Gravity, 20, S163, \dodoi{10.1088/0264-9381/20/10/319}

\bibitem[{{Cote} {et~al.}(2019){Cote}, {Abraham}, {Balogh}, {Capak}, {Carlberg}, {Cowan}, {Djazovski}, {Drissen}, {Drout}, {Dupuis}, {Evans}, {Fantin}, {Ferrarese}, {Fraser}, {Gallagher}, {Girard}, {Gleisinger}, {Grandmont}, {Hall}, {Hellmich}, {Hardy}, {Harrison}, {Hlozek}, {Haggard}, {Henault-Brunet}, {Hutchings}, {Khatu}, {Kavelaars}, {Laurin}, {Lavigne}, {Lisman}, {Marois}, {McCabe}, {Metchev}, {Moutard}, {Netterfield}, {Nikzad}, {Ouellette}, {Pass}, {Parker}, {Pazder}, {Percival}, {Rhodes}, {Robert}, {Rowe}, {Sanchez-Janssen}, {Sivakoff}, {Shapiro}, {Sawicki}, {Scott}, {Van Waerbeke}, \& {Venn}}]{Cote2019}
{Cote}, P., {Abraham}, B., {Balogh}, M., {et~al.} 2019, in Canadian Long Range Plan for Astronomy and Astrophysics White Papers, Vol. 2020, 18, \dodoi{10.5281/zenodo.3758463}

\bibitem[{{D'Antona} {et~al.}(2006){D'Antona}, {Ventura}, {Burderi}, \& {Teodorescu}}]{DAntona2006}
{D'Antona}, F., {Ventura}, P., {Burderi}, L., \& {Teodorescu}, A. 2006, \apj, 653, 1429, \dodoi{10.1086/507408}

\bibitem[{{de Miguel} {et~al.}(2018){de Miguel}, {Patterson}, {Kemp}, {Myers}, {Rea}, {Krajci}, {Monard}, \& {Cook}}]{deMiguel2018}
{de Miguel}, E., {Patterson}, J., {Kemp}, J., {et~al.} 2018, \apj, 852, 19, \dodoi{10.3847/1538-4357/aa9ed6}

\bibitem[{{Dekany} {et~al.}(2020){Dekany}, {Smith}, {Riddle}, {Feeney}, {Porter}, {Hale}, {Zolkower}, {Belicki}, {Kaye}, {Henning}, {Walters}, {Cromer}, {Delacroix}, {Rodriguez}, {Reiley}, {Mao}, {Hover}, {Murphy}, {Burruss}, {Baker}, {Kowalski}, {Reif}, {Mueller}, {Bellm}, {Graham}, \& {Kulkarni}}]{Dekany2020}
{Dekany}, R., {Smith}, R.~M., {Riddle}, R., {et~al.} 2020, \pasp, 132, 038001, \dodoi{10.1088/1538-3873/ab4ca2}

\bibitem[{{Deloye} {et~al.}(2007){Deloye}, {Taam}, {Winisdoerffer}, \& {Chabrier}}]{Deloye2007}
{Deloye}, C.~J., {Taam}, R.~E., {Winisdoerffer}, C., \& {Chabrier}, G. 2007, \mnras, 381, 525, \dodoi{10.1111/j.1365-2966.2007.12262.x}

\bibitem[{{Dhillon} {et~al.}(2007){Dhillon}, {Marsh}, {Stevenson}, {Atkinson}, {Kerry}, {Peacocke}, {Vick}, {Beard}, {Ives}, {Lunney}, {McLay}, {Tierney}, {Kelly}, {Littlefair}, {Nicholson}, {Pashley}, {Harlaftis}, \& {O'Brien}}]{Dhillon2007}
{Dhillon}, V.~S., {Marsh}, T.~R., {Stevenson}, M.~J., {et~al.} 2007, \mnras, 378, 825, \dodoi{10.1111/j.1365-2966.2007.11881.x}

\bibitem[{{Dhillon} {et~al.}(2021){Dhillon}, {Bezawada}, {Black}, {Dixon}, {Gamble}, {Gao}, {Henry}, {Kerry}, {Littlefair}, {Lunney}, {Marsh}, {Miller}, {Parsons}, {Ashley}, {Breedt}, {Brown}, {Dyer}, {Green}, {Pelisoli}, {Sahman}, {Wild}, {Ives}, {Mehrgan}, {Stegmeier}, {Dubbeldam}, {Morris}, {Osborn}, {Wilson}, {Casares}, {Mu{\~n}oz-Darias}, {Pall{\'e}}, {Rodr{\'\i}guez-Gil}, {Shahbaz}, {Torres}, {de Ugarte Postigo}, {Cabrera-Lavers}, {Corradi}, {Dom{\'\i}nguez}, \& {Garc{\'\i}a-Alvarez}}]{Dhillon2021}
{Dhillon}, V.~S., {Bezawada}, N., {Black}, M., {et~al.} 2021, \mnras, 507, 350, \dodoi{10.1093/mnras/stab2130}

\bibitem[{{Eggleton}(1983)}]{Eggleton1983}
{Eggleton}, P.~P. 1983, \apj, 268, 368, \dodoi{10.1086/160960}

\bibitem[{{Evans} {et~al.}(2023){Evans}, {Page}, {Beardmore}, {Eyles-Ferris}, {Osborne}, {Campana}, {Kennea}, \& {Cenko}}]{Evans2023}
{Evans}, P.~A., {Page}, K.~L., {Beardmore}, A.~P., {et~al.} 2023, \mnras, 518, 174, \dodoi{10.1093/mnras/stac2937}

\bibitem[{{Frank} {et~al.}(2002){Frank}, {King}, \& {Raine}}]{Frank2002}
{Frank}, J., {King}, A., \& {Raine}, D.~J. 2002, {Accretion Power in Astrophysics: Third Edition}

\bibitem[{{Gaia Collaboration} {et~al.}(2016){Gaia Collaboration}, {Prusti}, {de Bruijne}, {Brown}, {Vallenari}, {Babusiaux}, {Bailer-Jones}, {Bastian}, {Biermann}, {Evans}, {Eyer}, {Jansen}, {Jordi}, {Klioner}, {Lammers}, {Lindegren}, {Luri}, {Mignard}, {Milligan}, {Panem}, {Poinsignon}, {Pourbaix}, {Randich}, {Sarri}, {Sartoretti}, {Siddiqui}, {Soubiran}, {Valette}, {van Leeuwen}, {Walton}, {Aerts}, {Arenou}, {Cropper}, {Drimmel}, {H{\o}g}, {Katz}, {Lattanzi}, {O'Mullane}, {Grebel}, {Holland}, {Huc}, {Passot}, {Bramante}, {Cacciari}, {Casta{\~n}eda}, {Chaoul}, {Cheek}, {De Angeli}, {Fabricius}, {Guerra}, {Hern{\'a}ndez}, {Jean-Antoine-Piccolo}, {Masana}, {Messineo}, {Mowlavi}, {Nienartowicz}, {Ord{\'o}{\~n}ez-Blanco}, {Panuzzo}, {Portell}, {Richards}, {Riello}, {Seabroke}, {Tanga}, {Th{\'e}venin}, {Torra}, {Els}, {Gracia-Abril}, {Comoretto}, {Garcia-Reinaldos}, {Lock}, {Mercier}, {Altmann}, {Andrae}, {Astraatmadja}, {Bellas-Velidis}, {Benson}, {Berthier}, {Blomme}, {Busso}, {Carry}, {Cellino}, {Clementini},
  {Cowell}, {Creevey}, {Cuypers}, {Davidson}, {De Ridder}, {de Torres}, {Delchambre}, {Dell'Oro}, {Ducourant}, {Fr{\'e}mat}, {Garc{\'\i}a-Torres}, {Gosset}, {Halbwachs}, {Hambly}, {Harrison}, {Hauser}, {Hestroffer}, {Hodgkin}, {Huckle}, {Hutton}, {Jasniewicz}, {Jordan}, {Kontizas}, {Korn}, {Lanzafame}, {Manteiga}, {Moitinho}, {Muinonen}, {Osinde}, {Pancino}, {Pauwels}, {Petit}, {Recio-Blanco}, {Robin}, {Sarro}, {Siopis}, {Smith}, {Smith}, {Sozzetti}, {Thuillot}, {van Reeven}, {Viala}, {Abbas}, {Abreu Aramburu}, {Accart}, {Aguado}, {Allan}, {Allasia}, {Altavilla}, {{\'A}lvarez}, {Alves}, {Anderson}, {Andrei}, {Anglada Varela}, {Antiche}, {Antoja}, {Ant{\'o}n}, {Arcay}, {Atzei}, {Ayache}, {Bach}, {Baker}, {Balaguer-N{\'u}{\~n}ez}, {Barache}, {Barata}, {Barbier}, {Barblan}, {Baroni}, {Barrado y Navascu{\'e}s}, {Barros}, {Barstow}, {Becciani}, {Bellazzini}, {Bellei}, {Bello Garc{\'\i}a}, {Belokurov}, {Bendjoya}, {Berihuete}, {Bianchi}, {Bienaym{\'e}}, {Billebaud}, {Blagorodnova}, {Blanco-Cuaresma}, {Boch},
  {Bombrun}, {Borrachero}, {Bouquillon}, {Bourda}, {Bouy}, {Bragaglia}, {Breddels}, {Brouillet}, {Br{\"u}semeister}, {Bucciarelli}, {Budnik}, {Burgess}, {Burgon}, {Burlacu}, {Busonero}, {Buzzi}, {Caffau}, {Cambras}, {Campbell}, {Cancelliere}, {Cantat-Gaudin}, {Carlucci}, {Carrasco}, {Castellani}, {Charlot}, {Charnas}, {Charvet}, {Chassat}, {Chiavassa}, {Clotet}, {Cocozza}, {Collins}, {Collins}, {Costigan}, {Crifo}, {Cross}, {Crosta}, {Crowley}, {Dafonte}, {Damerdji}, {Dapergolas}, {David}, {David}, {De Cat}, {de Felice}, {de Laverny}, {De Luise}, {De March}, {de Martino}, {de Souza}, {Debosscher}, {del Pozo}, {Delbo}, {Delgado}, {Delgado}, {di Marco}, {Di Matteo}, {Diakite}, {Distefano}, {Dolding}, {Dos Anjos}, {Drazinos}, {Dur{\'a}n}, {Dzigan}, {Ecale}, {Edvardsson}, {Enke}, {Erdmann}, {Escolar}, {Espina}, {Evans}, {Eynard Bontemps}, {Fabre}, {Fabrizio}, {Faigler}, {Falc{\~a}o}, {Farr{\`a}s Casas}, {Faye}, {Federici}, {Fedorets}, {Fern{\'a}ndez-Hern{\'a}ndez}, {Fernique}, {Fienga}, {Figueras}, {Filippi},
  {Findeisen}, {Fonti}, {Fouesneau}, {Fraile}, {Fraser}, {Fuchs}, {Furnell}, {Gai}, {Galleti}, {Galluccio}, {Garabato}, {Garc{\'\i}a-Sedano}, {Gar{\'e}}, {Garofalo}, {Garralda}, {Gavras}, {Gerssen}, {Geyer}, {Gilmore}, {Girona}, {Giuffrida}, {Gomes}, {Gonz{\'a}lez-Marcos}, {Gonz{\'a}lez-N{\'u}{\~n}ez}, {Gonz{\'a}lez-Vidal}, {Granvik}, {Guerrier}, {Guillout}, {Guiraud}, {G{\'u}rpide}, {Guti{\'e}rrez-S{\'a}nchez}, {Guy}, {Haigron}, {Hatzidimitriou}, {Haywood}, {Heiter}, {Helmi}, {Hobbs}, {Hofmann}, {Holl}, {Holland}, {Hunt}, {Hypki}, {Icardi}, {Irwin}, {Jevardat de Fombelle}, {Jofr{\'e}}, {Jonker}, {Jorissen}, {Julbe}, {Karampelas}, {Kochoska}, {Kohley}, {Kolenberg}, {Kontizas}, {Koposov}, {Kordopatis}, {Koubsky}, {Kowalczyk}, {Krone-Martins}, {Kudryashova}, {Kull}, {Bachchan}, {Lacoste-Seris}, {Lanza}, {Lavigne}, {Le Poncin-Lafitte}, {Lebreton}, {Lebzelter}, {Leccia}, {Leclerc}, {Lecoeur-Taibi}, {Lemaitre}, {Lenhardt}, {Leroux}, {Liao}, {Licata}, {Lindstr{\o}m}, {Lister}, {Livanou}, {Lobel}, {L{\"o}ffler},
  {L{\'o}pez}, {Lopez-Lozano}, {Lorenz}, {Loureiro}, {MacDonald}, {Magalh{\~a}es Fernandes}, {Managau}, {Mann}, {Mantelet}, {Marchal}, {Marchant}, {Marconi}, {Marie}, {Marinoni}, {Marrese}, {Marschalk{\'o}}, {Marshall}, {Mart{\'\i}n-Fleitas}, {Martino}, {Mary}, {Matijevi{\v{c}}}, {Mazeh}, {McMillan}, {Messina}, {Mestre}, {Michalik}, {Millar}, {Miranda}, {Molina}, {Molinaro}, {Molinaro}, {Moln{\'a}r}, {Moniez}, {Montegriffo}, {Monteiro}, {Mor}, {Mora}, {Morbidelli}, {Morel}, {Morgenthaler}, {Morley}, {Morris}, {Mulone}, {Muraveva}, {Musella}, {Narbonne}, {Nelemans}, {Nicastro}, {Noval}, {Ord{\'e}novic}, {Ordieres-Mer{\'e}}, {Osborne}, {Pagani}, {Pagano}, {Pailler}, {Palacin}, {Palaversa}, {Parsons}, {Paulsen}, {Pecoraro}, {Pedrosa}, {Pentik{\"a}inen}, {Pereira}, {Pichon}, {Piersimoni}, {Pineau}, {Plachy}, {Plum}, {Poujoulet}, {Pr{\v{s}}a}, {Pulone}, {Ragaini}, {Rago}, {Rambaux}, {Ramos-Lerate}, {Ranalli}, {Rauw}, {Read}, {Regibo}, {Renk}, {Reyl{\'e}}, {Ribeiro}, {Rimoldini}, {Ripepi}, {Riva}, {Rixon},
  {Roelens}, {Romero-G{\'o}mez}, {Rowell}, {Royer}, {Rudolph}, {Ruiz-Dern}, {Sadowski}, {Sagrist{\`a} Sell{\'e}s}, {Sahlmann}, {Salgado}, {Salguero}, {Sarasso}, {Savietto}, {Schnorhk}, {Schultheis}, {Sciacca}, {Segol}, {Segovia}, {Segransan}, {Serpell}, {Shih}, {Smareglia}, {Smart}, {Smith}, {Solano}, {Solitro}, {Sordo}, {Soria Nieto}, {Souchay}, {Spagna}, {Spoto}, {Stampa}, {Steele}, {Steidelm{\"u}ller}, {Stephenson}, {Stoev}, {Suess}, {S{\"u}veges}, {Surdej}, {Szabados}, {Szegedi-Elek}, {Tapiador}, {Taris}, {Tauran}, {Taylor}, {Teixeira}, {Terrett}, {Tingley}, {Trager}, {Turon}, {Ulla}, {Utrilla}, {Valentini}, {van Elteren}, {Van Hemelryck}, {van Leeuwen}, {Varadi}, {Vecchiato}, {Veljanoski}, {Via}, {Vicente}, {Vogt}, {Voss}, {Votruba}, {Voutsinas}, {Walmsley}, {Weiler}, {Weingrill}, {Werner}, {Wevers}, {Whitehead}, {Wyrzykowski}, {Yoldas}, {{\v{Z}}erjal}, {Zucker}, {Zurbach}, {Zwitter}, {Alecu}, {Allen}, {Allende Prieto}, {Amorim}, {Anglada-Escud{\'e}}, {Arsenijevic}, {Azaz}, {Balm}, {Beck}, {Bernstein},
  {Bigot}, {Bijaoui}, {Blasco}, {Bonfigli}, {Bono}, {Boudreault}, {Bressan}, {Brown}, {Brunet}, {Bunclark}, {Buonanno}, {Butkevich}, {Carret}, {Carrion}, {Chemin}, {Ch{\'e}reau}, {Corcione}, {Darmigny}, {de Boer}, {de Teodoro}, {de Zeeuw}, {Delle Luche}, {Domingues}, {Dubath}, {Fodor}, {Fr{\'e}zouls}, {Fries}, {Fustes}, {Fyfe}, {Gallardo}, {Gallegos}, {Gardiol}, {Gebran}, {Gomboc}, {G{\'o}mez}, {Grux}, {Gueguen}, {Heyrovsky}, {Hoar}, {Iannicola}, {Isasi Parache}, {Janotto}, {Joliet}, {Jonckheere}, {Keil}, {Kim}, {Klagyivik}, {Klar}, {Knude}, {Kochukhov}, {Kolka}, {Kos}, {Kutka}, {Lainey}, {LeBouquin}, {Liu}, {Loreggia}, {Makarov}, {Marseille}, {Martayan}, {Martinez-Rubi}, {Massart}, {Meynadier}, {Mignot}, {Munari}, {Nguyen}, {Nordlander}, {Ocvirk}, {O'Flaherty}, {Olias Sanz}, {Ortiz}, {Osorio}, {Oszkiewicz}, {Ouzounis}, {Palmer}, {Park}, {Pasquato}, {Peltzer}, {Peralta}, {P{\'e}turaud}, {Pieniluoma}, {Pigozzi}, {Poels}, {Prat}, {Prod'homme}, {Raison}, {Rebordao}, {Risquez}, {Rocca-Volmerange}, {Rosen},
  {Ruiz-Fuertes}, {Russo}, {Sembay}, {Serraller Vizcaino}, {Short}, {Siebert}, {Silva}, {Sinachopoulos}, {Slezak}, {Soffel}, {Sosnowska}, {Strai{\v{z}}ys}, {ter Linden}, {Terrell}, {Theil}, {Tiede}, {Troisi}, {Tsalmantza}, {Tur}, {Vaccari}, {Vachier}, {Valles}, {Van Hamme}, {Veltz}, {Virtanen}, {Wallut}, {Wichmann}, {Wilkinson}, {Ziaeepour}, \& {Zschocke}}]{Gaia2016}
{Gaia Collaboration}, {Prusti}, T., {de Bruijne}, J.~H.~J., {et~al.} 2016, \aap, 595, A1, \dodoi{10.1051/0004-6361/201629272}

\bibitem[{{G{\"a}nsicke} {et~al.}(2003){G{\"a}nsicke}, {Szkody}, {de Martino}, {Beuermann}, {Long}, {Sion}, {Knigge}, {Marsh}, \& {Hubeny}}]{Gansicke2003}
{G{\"a}nsicke}, B.~T., {Szkody}, P., {de Martino}, D., {et~al.} 2003, \apj, 594, 443, \dodoi{10.1086/376902}

\bibitem[{{Gehrels} {et~al.}(2004){Gehrels}, {Chincarini}, {Giommi}, {Mason}, {Nousek}, {Wells}, {White}, {Barthelmy}, {Burrows}, {Cominsky}, {Hurley}, {Marshall}, {M{\'e}sz{\'a}ros}, {Roming}, {Angelini}, {Barbier}, {Belloni}, {Campana}, {Caraveo}, {Chester}, {Citterio}, {Cline}, {Cropper}, {Cummings}, {Dean}, {Feigelson}, {Fenimore}, {Frail}, {Fruchter}, {Garmire}, {Gendreau}, {Ghisellini}, {Greiner}, {Hill}, {Hunsberger}, {Krimm}, {Kulkarni}, {Kumar}, {Lebrun}, {Lloyd-Ronning}, {Markwardt}, {Mattson}, {Mushotzky}, {Norris}, {Osborne}, {Paczynski}, {Palmer}, {Park}, {Parsons}, {Paul}, {Rees}, {Reynolds}, {Rhoads}, {Sasseen}, {Schaefer}, {Short}, {Smale}, {Smith}, {Stella}, {Tagliaferri}, {Takahashi}, {Tashiro}, {Townsley}, {Tueller}, {Turner}, {Vietri}, {Voges}, {Ward}, {Willingale}, {Zerbi}, \& {Zhang}}]{Gehrels2004}
{Gehrels}, N., {Chincarini}, G., {Giommi}, P., {et~al.} 2004, \apj, 611, 1005, \dodoi{10.1086/422091}

\bibitem[{{Gokhale} {et~al.}(2007){Gokhale}, {Peng}, \& {Frank}}]{Gokhale2007}
{Gokhale}, V., {Peng}, X.~M., \& {Frank}, J. 2007, \apj, 655, 1010, \dodoi{10.1086/510119}

\bibitem[{{Graham} {et~al.}(2019){Graham}, {Kulkarni}, {Bellm}, {Adams}, {Barbarino}, {Blagorodnova}, {Bodewits}, {Bolin}, {Brady}, {Cenko}, {Chang}, {Coughlin}, {De}, {Eadie}, {Farnham}, {Feindt}, {Franckowiak}, {Fremling}, {Gezari}, {Ghosh}, {Goldstein}, {Golkhou}, {Goobar}, {Ho}, {Huppenkothen}, {Ivezi{\'c}}, {Jones}, {Juric}, {Kaplan}, {Kasliwal}, {Kelley}, {Kupfer}, {Lee}, {Lin}, {Lunnan}, {Mahabal}, {Miller}, {Ngeow}, {Nugent}, {Ofek}, {Prince}, {Rauch}, {van Roestel}, {Schulze}, {Singer}, {Sollerman}, {Taddia}, {Yan}, {Ye}, {Yu}, {Barlow}, {Bauer}, {Beck}, {Belicki}, {Biswas}, {Brinnel}, {Brooke}, {Bue}, {Bulla}, {Burruss}, {Connolly}, {Cromer}, {Cunningham}, {Dekany}, {Delacroix}, {Desai}, {Duev}, {Feeney}, {Flynn}, {Frederick}, {Gal-Yam}, {Giomi}, {Groom}, {Hacopians}, {Hale}, {Helou}, {Henning}, {Hover}, {Hillenbrand}, {Howell}, {Hung}, {Imel}, {Ip}, {Jackson}, {Kaspi}, {Kaye}, {Kowalski}, {Kramer}, {Kuhn}, {Landry}, {Laher}, {Mao}, {Masci}, {Monkewitz}, {Murphy}, {Nordin}, {Patterson}, {Penprase},
  {Porter}, {Rebbapragada}, {Reiley}, {Riddle}, {Rigault}, {Rodriguez}, {Rusholme}, {van Santen}, {Shupe}, {Smith}, {Soumagnac}, {Stein}, {Surace}, {Szkody}, {Terek}, {Van Sistine}, {van Velzen}, {Vestrand}, {Walters}, {Ward}, {Zhang}, \& {Zolkower}}]{Graham2019}
{Graham}, M.~J., {Kulkarni}, S.~R., {Bellm}, E.~C., {et~al.} 2019, \pasp, 131, 078001, \dodoi{10.1088/1538-3873/ab006c}

\bibitem[{{Green}(2024)}]{Green2024}
{Green}, M. 2024, \dodoi{10.5281/zenodo.8276712}

\bibitem[{{Green} {et~al.}(2018{\natexlab{a}}){Green}, {Hermes}, {Marsh}, {Steeghs}, {Bell}, {Littlefair}, {Parsons}, {Dennihy}, {Fuchs}, {Reding}, {Kaiser}, {Ashley}, {Breedt}, {Dhillon}, {Gentile Fusillo}, {Kerry}, \& {Sahman}}]{Green2018b}
{Green}, M.~J., {Hermes}, J.~J., {Marsh}, T.~R., {et~al.} 2018{\natexlab{a}}, \mnras, 477, 5646, \dodoi{10.1093/mnras/sty1032}

\bibitem[{{Green} {et~al.}(2018{\natexlab{b}}){Green}, {Marsh}, {Steeghs}, {Kupfer}, {Ashley}, {Bloemen}, {Breedt}, {Campbell}, {Chakpor}, {Copperwheat}, {Dhillon}, {Hallinan}, {Hardy}, {Hermes}, {Kerry}, {Littlefair}, {Milburn}, {Parsons}, {Prasert}, {van Roestel}, {Sahman}, \& {Singh}}]{Green2018a}
{Green}, M.~J., {Marsh}, T.~R., {Steeghs}, D.~T.~H., {et~al.} 2018{\natexlab{b}}, \mnras, 476, 1663, \dodoi{10.1093/mnras/sty299}

\bibitem[{{Harding} {et~al.}(2016){Harding}, {Hallinan}, {Milburn}, {Gardner}, {Konidaris}, {Singh}, {Shao}, {Sandhu}, {Kyne}, \& {Schlichting}}]{Harding2016}
{Harding}, L.~K., {Hallinan}, G., {Milburn}, J., {et~al.} 2016, \mnras, 457, 3036, \dodoi{10.1093/mnras/stw094}

\bibitem[{{Heber}(2016)}]{Heber2016}
{Heber}, U. 2016, \pasp, 128, 082001, \dodoi{10.1088/1538-3873/128/966/082001}

\bibitem[{{Hermes} {et~al.}(2012){Hermes}, {Kilic}, {Brown}, {Winget}, {Allende Prieto}, {Gianninas}, {Mukadam}, {Cabrera-Lavers}, \& {Kenyon}}]{Hermes2012}
{Hermes}, J.~J., {Kilic}, M., {Brown}, W.~R., {et~al.} 2012, \apjl, 757, L21, \dodoi{10.1088/2041-8205/757/2/L21}

\bibitem[{{Hillebrandt} {et~al.}(2013){Hillebrandt}, {Kromer}, {R{\"o}pke}, \& {Ruiter}}]{Hillebrandt2013}
{Hillebrandt}, W., {Kromer}, M., {R{\"o}pke}, F.~K., \& {Ruiter}, A.~J. 2013, Frontiers of Physics, 8, 116, \dodoi{10.1007/s11467-013-0303-2}

\bibitem[{{Israel} {et~al.}(1999){Israel}, {Panzera}, {Campana}, {Lazzati}, {Covino}, {Tagliaferri}, \& {Stella}}]{Israel1999}
{Israel}, G.~L., {Panzera}, M.~R., {Campana}, S., {et~al.} 1999, \aap, 349, L1

\bibitem[{{Jha} {et~al.}(2019){Jha}, {Maguire}, \& {Sullivan}}]{Jha2019}
{Jha}, S.~W., {Maguire}, K., \& {Sullivan}, M. 2019, Nature Astronomy, 3, 706, \dodoi{10.1038/s41550-019-0858-0}

\bibitem[{{Kalomeni} {et~al.}(2016){Kalomeni}, {Nelson}, {Rappaport}, {Molnar}, {Quintin}, \& {Yakut}}]{Kalomeni2016}
{Kalomeni}, B., {Nelson}, L., {Rappaport}, S., {et~al.} 2016, \apj, 833, 83, \dodoi{10.3847/1538-4357/833/1/83}

\bibitem[{{Kaplan} {et~al.}(2012){Kaplan}, {Bildsten}, \& {Steinfadt}}]{Kaplan2012}
{Kaplan}, D.~L., {Bildsten}, L., \& {Steinfadt}, J. D.~R. 2012, \apj, 758, 64, \dodoi{10.1088/0004-637X/758/1/64}

\bibitem[{{Kondo} {et~al.}(1984){Kondo}, {Noguchi}, \& {Maehara}}]{Konda1984}
{Kondo}, M., {Noguchi}, T., \& {Maehara}, H. 1984, Annals of the Tokyo Astronomical Observatory, 20, 130

\bibitem[{{Kremer} {et~al.}(2015){Kremer}, {Sepinsky}, \& {Kalogera}}]{Kremer2015}
{Kremer}, K., {Sepinsky}, J., \& {Kalogera}, V. 2015, \apj, 806, 76, \dodoi{10.1088/0004-637X/806/1/76}

\bibitem[{{Kulkarni} {et~al.}(2021){Kulkarni}, {Harrison}, {Grefenstette}, {Earnshaw}, {Andreoni}, {Berg}, {Bloom}, {Cenko}, {Chornock}, {Christiansen}, {Coughlin}, {Wuollet Criswell}, {Darvish}, {Das}, {De}, {Dessart}, {Dixon}, {Dorsman}, {El-Badry}, {Evans}, {Ford}, {Fremling}, {Gansicke}, {Gezari}, {Goetberg}, {Green}, {Graham}, {Heida}, {Ho}, {Jaodand}, {Johns-Krull}, {Kasliwal}, {Lazzarini}, {Lu}, {Margutti}, {Martin}, {Masters}, {McKernan}, {Naze}, {Nissanke}, {Parazin}, {Perley}, {Phinney}, {Piro}, {Raaijmakers}, {Rauw}, {Rodriguez}, {Sana}, {Senchyna}, {Singer}, {Spake}, {Stassun}, {Stern}, {Teplitz}, {Weisz}, \& {Yao}}]{Kulkarni2021}
{Kulkarni}, S.~R., {Harrison}, F.~A., {Grefenstette}, B.~W., {et~al.} 2021, arXiv e-prints, arXiv:2111.15608, \dodoi{10.48550/arXiv.2111.15608}

\bibitem[{{Kupfer} {et~al.}(2018){Kupfer}, {Korol}, {Shah}, {Nelemans}, {Marsh}, {Ramsay}, {Groot}, {Steeghs}, \& {Rossi}}]{Kupfer2018}
{Kupfer}, T., {Korol}, V., {Shah}, S., {et~al.} 2018, \mnras, 480, 302, \dodoi{10.1093/mnras/sty1545}

\bibitem[{{Kupfer} {et~al.}(2024){Kupfer}, {Korol}, {Littenberg}, {Shah}, {Savalle}, {Groot}, {Marsh}, {Le Jeune}, {Nelemans}, {Pala}, {Petiteau}, {Ramsay}, {Steeghs}, \& {Babak}}]{Kupfer2024}
{Kupfer}, T., {Korol}, V., {Littenberg}, T.~B., {et~al.} 2024, \apj, 963, 100, \dodoi{10.3847/1538-4357/ad2068}

\bibitem[{{Levitan} {et~al.}(2015){Levitan}, {Groot}, {Prince}, {Kulkarni}, {Laher}, {Ofek}, {Sesar}, \& {Surace}}]{Levitan2015}
{Levitan}, D., {Groot}, P.~J., {Prince}, T.~A., {et~al.} 2015, \mnras, 446, 391, \dodoi{10.1093/mnras/stu2105}

\bibitem[{{Littenberg} {et~al.}(2020){Littenberg}, {Cornish}, {Lackeos}, \& {Robson}}]{Littenberg2020}
{Littenberg}, T.~B., {Cornish}, N.~J., {Lackeos}, K., \& {Robson}, T. 2020, \prd, 101, 123021, \dodoi{10.1103/PhysRevD.101.123021}

\bibitem[{{Liu} {et~al.}(2023){Liu}, {R{\"o}pke}, \& {Han}}]{Liu2023}
{Liu}, Z.-W., {R{\"o}pke}, F.~K., \& {Han}, Z. 2023, Research in Astronomy and Astrophysics, 23, 082001, \dodoi{10.1088/1674-4527/acd89e}

\bibitem[{{Luo} {et~al.}(2016){Luo}, {Chen}, {Duan}, {Gong}, {Hu}, {Ji}, {Liu}, {Mei}, {Milyukov}, {Sazhin}, {Shao}, {Toth}, {Tu}, {Wang}, {Wang}, {Yeh}, {Zhan}, {Zhang}, {Zharov}, \& {Zhou}}]{Luo2016}
{Luo}, J., {Chen}, L.-S., {Duan}, H.-Z., {et~al.} 2016, Classical and Quantum Gravity, 33, 035010, \dodoi{10.1088/0264-9381/33/3/035010}

\bibitem[{{Maitra} {et~al.}(2024){Maitra}, {Haberl}, {Vasilopoulos}, {Rau}, {Schwope}, {Friedrich}, {Buckley}, {Valdes}, {Lang}, \& {Macfarlane}}]{Maitra2024}
{Maitra}, C., {Haberl}, F., {Vasilopoulos}, G., {et~al.} 2024, \aap, 683, A21, \dodoi{10.1051/0004-6361/202347811}

\bibitem[{{Maoz} {et~al.}(2014){Maoz}, {Mannucci}, \& {Nelemans}}]{Maoz2014}
{Maoz}, D., {Mannucci}, F., \& {Nelemans}, G. 2014, \araa, 52, 107, \dodoi{10.1146/annurev-astro-082812-141031}

\bibitem[{{Marsh} {et~al.}(2004){Marsh}, {Nelemans}, \& {Steeghs}}]{Marsh2004}
{Marsh}, T.~R., {Nelemans}, G., \& {Steeghs}, D. 2004, \mnras, 350, 113, \dodoi{10.1111/j.1365-2966.2004.07564.x}

\bibitem[{{Marsh} \& {Steeghs}(2002)}]{Marsh2002}
{Marsh}, T.~R., \& {Steeghs}, D. 2002, \mnras, 331, L7, \dodoi{10.1046/j.1365-8711.2002.05346.x}

\bibitem[{{Marsh} {et~al.}(1995){Marsh}, {Wood}, {Horne}, \& {Lambert}}]{Marsh1995}
{Marsh}, T.~R., {Wood}, J.~H., {Horne}, K., \& {Lambert}, D. 1995, \mnras, 274, 452, \dodoi{10.1093/mnras/274.2.452}

\bibitem[{{Masci} {et~al.}(2019){Masci}, {Laher}, {Rusholme}, {Shupe}, {Groom}, {Surace}, {Jackson}, {Monkewitz}, {Beck}, {Flynn}, {Terek}, {Landry}, {Hacopians}, {Desai}, {Howell}, {Brooke}, {Imel}, {Wachter}, {Ye}, {Lin}, {Cenko}, {Cunningham}, {Rebbapragada}, {Bue}, {Miller}, {Mahabal}, {Bellm}, {Patterson}, {Juri{\'c}}, {Golkhou}, {Ofek}, {Walters}, {Graham}, {Kasliwal}, {Dekany}, {Kupfer}, {Burdge}, {Cannella}, {Barlow}, {Van Sistine}, {Giomi}, {Fremling}, {Blagorodnova}, {Levitan}, {Riddle}, {Smith}, {Helou}, {Prince}, \& {Kulkarni}}]{Masci2019}
{Masci}, F.~J., {Laher}, R.~R., {Rusholme}, B., {et~al.} 2019, \pasp, 131, 018003, \dodoi{10.1088/1538-3873/aae8ac}

\bibitem[{{Motch} {et~al.}(1996){Motch}, {Haberl}, {Guillout}, {Pakull}, {Reinsch}, \& {Krautter}}]{Motch1996}
{Motch}, C., {Haberl}, F., {Guillout}, P., {et~al.} 1996, \aap, 307, 459

\bibitem[{{Motl} {et~al.}(2007){Motl}, {Frank}, {Tohline}, \& {D'Souza}}]{Motl2007}
{Motl}, P.~M., {Frank}, J., {Tohline}, J.~E., \& {D'Souza}, M. C.~R. 2007, \apj, 670, 1314, \dodoi{10.1086/522076}

\bibitem[{{Munday}(in prep.)}]{MundayInPrep}
{Munday}, J. in prep.

\bibitem[{{Munday} {et~al.}(2023){Munday}, {Marsh}, {Hollands}, {Pelisoli}, {Steeghs}, {Hakala}, {Breedt}, {Brown}, {Dhillon}, {Dyer}, {Green}, {Kerry}, {Littlefair}, {Parsons}, {Sahman}, {Somjit}, {Sukaum}, \& {Wild}}]{Munday2023}
{Munday}, J., {Marsh}, T.~R., {Hollands}, M., {et~al.} 2023, \mnras, 518, 5123, \dodoi{10.1093/mnras/stac3385}

\bibitem[{{Nelemans}(2005)}]{Nelemans2005}
{Nelemans}, G. 2005, in Astronomical Society of the Pacific Conference Series, Vol. 330, The Astrophysics of Cataclysmic Variables and Related Objects, ed. J.~M. {Hameury} \& J.~P. {Lasota}, 27, \dodoi{10.48550/arXiv.astro-ph/0409676}

\bibitem[{{Nelemans} {et~al.}(2001){Nelemans}, {Portegies Zwart}, {Verbunt}, \& {Yungelson}}]{Nelemans2001}
{Nelemans}, G., {Portegies Zwart}, S.~F., {Verbunt}, F., \& {Yungelson}, L.~R. 2001, \aap, 368, 939, \dodoi{10.1051/0004-6361:20010049}

\bibitem[{{Nelemans} {et~al.}(2004){Nelemans}, {Yungelson}, \& {Portegies Zwart}}]{Nelemans2004}
{Nelemans}, G., {Yungelson}, L.~R., \& {Portegies Zwart}, S.~F. 2004, \mnras, 349, 181, \dodoi{10.1111/j.1365-2966.2004.07479.x}

\bibitem[{{Nelemans} {et~al.}(2010){Nelemans}, {Yungelson}, {van der Sluys}, \& {Tout}}]{Nelemans2010}
{Nelemans}, G., {Yungelson}, L.~R., {van der Sluys}, M.~V., \& {Tout}, C.~A. 2010, \mnras, 401, 1347, \dodoi{10.1111/j.1365-2966.2009.15731.x}

\bibitem[{{Oke} {et~al.}(1995){Oke}, {Cohen}, {Carr}, {Cromer}, {Dingizian}, {Harris}, {Labrecque}, {Lucinio}, {Schaal}, {Epps}, \& {Miller}}]{Oke1995}
{Oke}, J.~B., {Cohen}, J.~G., {Carr}, M., {et~al.} 1995, \pasp, 107, 375, \dodoi{10.1086/133562}

\bibitem[{{Paczy{\'n}ski}(1967)}]{Paczynski1967}
{Paczy{\'n}ski}, B. 1967, \actaa, 17, 287

\bibitem[{{Paczy{\'n}ski}(1971)}]{Paczynski1971}
---. 1971, \araa, 9, 183, \dodoi{10.1146/annurev.aa.09.090171.001151}

\bibitem[{{Paxton} {et~al.}(2011){Paxton}, {Bildsten}, {Dotter}, {Herwig}, {Lesaffre}, \& {Timmes}}]{Paxton2011}
{Paxton}, B., {Bildsten}, L., {Dotter}, A., {et~al.} 2011, \apjs, 192, 3, \dodoi{10.1088/0067-0049/192/1/3}

\bibitem[{{Paxton} {et~al.}(2013){Paxton}, {Cantiello}, {Arras}, {Bildsten}, {Brown}, {Dotter}, {Mankovich}, {Montgomery}, {Stello}, {Timmes}, \& {Townsend}}]{Paxton2013}
{Paxton}, B., {Cantiello}, M., {Arras}, P., {et~al.} 2013, \apjs, 208, 4, \dodoi{10.1088/0067-0049/208/1/4}

\bibitem[{{Paxton} {et~al.}(2015){Paxton}, {Marchant}, {Schwab}, {Bauer}, {Bildsten}, {Cantiello}, {Dessart}, {Farmer}, {Hu}, {Langer}, {Townsend}, {Townsley}, \& {Timmes}}]{Paxton2015}
{Paxton}, B., {Marchant}, P., {Schwab}, J., {et~al.} 2015, \apjs, 220, 15, \dodoi{10.1088/0067-0049/220/1/15}

\bibitem[{{Paxton} {et~al.}(2018){Paxton}, {Schwab}, {Bauer}, {Bildsten}, {Blinnikov}, {Duffell}, {Farmer}, {Goldberg}, {Marchant}, {Sorokina}, {Thoul}, {Townsend}, \& {Timmes}}]{Paxton2018}
{Paxton}, B., {Schwab}, J., {Bauer}, E.~B., {et~al.} 2018, \apjs, 234, 34, \dodoi{10.3847/1538-4365/aaa5a8}

\bibitem[{{Paxton} {et~al.}(2019){Paxton}, {Smolec}, {Schwab}, {Gautschy}, {Bildsten}, {Cantiello}, {Dotter}, {Farmer}, {Goldberg}, {Jermyn}, {Kanbur}, {Marchant}, {Thoul}, {Townsend}, {Wolf}, {Zhang}, \& {Timmes}}]{Paxton2019}
{Paxton}, B., {Smolec}, R., {Schwab}, J., {et~al.} 2019, \apjs, 243, 10, \dodoi{10.3847/1538-4365/ab2241}

\bibitem[{{Perley}(2019)}]{Perley2019}
{Perley}, D.~A. 2019, \pasp, 131, 084503, \dodoi{10.1088/1538-3873/ab215d}

\bibitem[{{Peters}(1964)}]{Peters1964}
{Peters}, P.~C. 1964, Physical Review, 136, 1224, \dodoi{10.1103/PhysRev.136.B1224}

\bibitem[{{Ramsay} {et~al.}(2018){Ramsay}, {Green}, {Marsh}, {Kupfer}, {Breedt}, {Korol}, {Groot}, {Knigge}, {Nelemans}, {Steeghs}, {Woudt}, \& {Aungwerojwit}}]{Ramsay2018}
{Ramsay}, G., {Green}, M.~J., {Marsh}, T.~R., {et~al.} 2018, \aap, 620, A141, \dodoi{10.1051/0004-6361/201834261}

\bibitem[{{Rau} {et~al.}(2010){Rau}, {Roelofs}, {Groot}, {Marsh}, {Nelemans}, {Steeghs}, {Salvato}, \& {Kasliwal}}]{Rau2010}
{Rau}, A., {Roelofs}, G.~H.~A., {Groot}, P.~J., {et~al.} 2010, \apj, 708, 456, \dodoi{10.1088/0004-637X/708/1/456}

\bibitem[{{Rodriguez} {et~al.}(2023){Rodriguez}, {Galiullin}, {Gilfanov}, {Kulkarni}, {Khamitov}, {Bikmaev}, {van Roestel}, {Yungelson}, {El-Badry}, {Sunayev}, {Prince}, {Buntov}, {Caiazzo}, {Drake}, {Gorbachev}, {Graham}, {Gumerov}, {Irtuganov}, {Laher}, {Masci}, {Medvedev}, {Purdum}, {Sakhibullin}, {Sklyanov}, {Smith}, {Szkody}, \& {Vanderbosch}}]{Rodriguez2023}
{Rodriguez}, A.~C., {Galiullin}, I., {Gilfanov}, M., {et~al.} 2023, \apj, 954, 63, \dodoi{10.3847/1538-4357/ace698}

\bibitem[{{Rodriguez} {et~al.}(2024){Rodriguez}, {El-Badry}, {Suleimanov}, {Pala}, {Kulkarni}, {Gaensicke}, {Mori}, {Rich}, {Sarkar}, {Bao}, {Lopes de Oliveira}, {Ramsay}, {Szkody}, {Graham}, {Prince}, {Caiazzo}, {Vanderbosch}, {van Roestel}, {Das}, {Qin}, {Kasliwal}, {Wold}, {Groom}, {Reiley}, \& {Riddle}}]{Rodriguez2024}
{Rodriguez}, A.~C., {El-Badry}, K., {Suleimanov}, V., {et~al.} 2024, arXiv e-prints, arXiv:2408.16053, \dodoi{10.48550/arXiv.2408.16053}

\bibitem[{{Roelofs} {et~al.}(2007){Roelofs}, {Nelemans}, \& {Groot}}]{Roelofs2007}
{Roelofs}, G.~H.~A., {Nelemans}, G., \& {Groot}, P.~J. 2007, \mnras, 382, 685, \dodoi{10.1111/j.1365-2966.2007.12451.x}

\bibitem[{{Roelofs} {et~al.}(2010){Roelofs}, {Rau}, {Marsh}, {Steeghs}, {Groot}, \& {Nelemans}}]{Roelofs2010}
{Roelofs}, G. H.~A., {Rau}, A., {Marsh}, T.~R., {et~al.} 2010, \apjl, 711, L138, \dodoi{10.1088/2041-8205/711/2/L138}

\bibitem[{{Roelofs} {et~al.}(2009){Roelofs}, {Groot}, {Steeghs}, {Rau}, {de Groot}, {Marsh}, {Nelemans}, {Liebert}, \& {Woudt}}]{Roelofs2009}
{Roelofs}, G.~H.~A., {Groot}, P.~J., {Steeghs}, D., {et~al.} 2009, \mnras, 394, 367, \dodoi{10.1111/j.1365-2966.2008.14288.x}

\bibitem[{{Ruan} {et~al.}(2020){Ruan}, {Guo}, {Cai}, \& {Zhang}}]{Ruan2020}
{Ruan}, W.-H., {Guo}, Z.-K., {Cai}, R.-G., \& {Zhang}, Y.-Z. 2020, International Journal of Modern Physics A, 35, 2050075, \dodoi{10.1142/S0217751X2050075X}

\bibitem[{{Ruiter} {et~al.}(2009){Ruiter}, {Belczynski}, \& {Fryer}}]{Ruiter2009}
{Ruiter}, A.~J., {Belczynski}, K., \& {Fryer}, C. 2009, \apj, 699, 2026, \dodoi{10.1088/0004-637X/699/2/2026}

\bibitem[{{Savonije} {et~al.}(1986){Savonije}, {de Kool}, \& {van den Heuvel}}]{Savonije1986}
{Savonije}, G.~J., {de Kool}, M., \& {van den Heuvel}, E.~P.~J. 1986, \aap, 155, 51

\bibitem[{{Shah} {et~al.}(2012){Shah}, {van der Sluys}, \& {Nelemans}}]{Shah2012}
{Shah}, S., {van der Sluys}, M., \& {Nelemans}, G. 2012, \aap, 544, A153, \dodoi{10.1051/0004-6361/201219309}

\bibitem[{{Shen}(2015)}]{Shen2015}
{Shen}, K.~J. 2015, \apjl, 805, L6, \dodoi{10.1088/2041-8205/805/1/L6}

\bibitem[{{Shen} \& {Moore}(2014)}]{Shen2014}
{Shen}, K.~J., \& {Moore}, K. 2014, \apj, 797, 46, \dodoi{10.1088/0004-637X/797/1/46}

\bibitem[{{Smak}(1967)}]{Smak1967}
{Smak}, J. 1967, \actaa, 17, 255

\bibitem[{{Smak}(1985)}]{Smak1985}
---. 1985, \actaa, 35, 351

\bibitem[{{Solheim}(2010)}]{Solheim2010}
{Solheim}, J.~E. 2010, \pasp, 122, 1133, \dodoi{10.1086/656680}

\bibitem[{{Strohmayer}(2004{\natexlab{a}})}]{Strohmayer2004b}
{Strohmayer}, T.~E. 2004{\natexlab{a}}, \apj, 614, 358, \dodoi{10.1086/423615}

\bibitem[{{Strohmayer}(2004{\natexlab{b}})}]{Strohmayer2004a}
---. 2004{\natexlab{b}}, \apj, 610, 416, \dodoi{10.1086/421384}

\bibitem[{{Strohmayer}(2021)}]{Strohmayer2021}
---. 2021, \apjl, 912, L8, \dodoi{10.3847/2041-8213/abf3cc}

\bibitem[{{Toubiana} {et~al.}(2024){Toubiana}, {Karnesis}, {Lamberts}, \& {Miller}}]{Toubiana2024}
{Toubiana}, A., {Karnesis}, N., {Lamberts}, A., \& {Miller}, M.~C. 2024, arXiv e-prints, arXiv:2403.16867, \dodoi{10.48550/arXiv.2403.16867}

\bibitem[{{van Roestel} {et~al.}(2022){van Roestel}, {Kupfer}, {Green}, {Wong}, {Bildsten}, {Burdge}, {Prince}, {Marsh}, {Szkody}, {Fremling}, {Graham}, {Dhillon}, {Littlefair}, {Bellm}, {Coughlin}, {Duev}, {Goldstein}, {Laher}, {Rusholme}, {Riddle}, {Dekany}, \& {Kulkarni}}]{vanRoestel2022}
{van Roestel}, J., {Kupfer}, T., {Green}, M.~J., {et~al.} 2022, \mnras, 512, 5440, \dodoi{10.1093/mnras/stab2421}

\bibitem[{{Verbunt} \& {Rappaport}(1988)}]{Verbunt1988}
{Verbunt}, F., \& {Rappaport}, S. 1988, \apj, 332, 193, \dodoi{10.1086/166645}

\bibitem[{{Warner}(1995)}]{Warner1995}
{Warner}, B. 1995, \apss, 225, 249, \dodoi{10.1007/BF00613240}

\bibitem[{{Wong} \& {Bildsten}(2021)}]{Wong2021}
{Wong}, T. L.~S., \& {Bildsten}, L. 2021, \apj, 923, 125, \dodoi{10.3847/1538-4357/ac2b2a}

\bibitem[{{Yungelson}(2008)}]{Yungelson2008}
{Yungelson}, L.~R. 2008, Astronomy Letters, 34, 620, \dodoi{10.1134/S1063773708090053}

\end{thebibliography}
\bibliographystyle{aasjournal}

\end{document}